\newcommand \Mpc {h^{-1}{\rm Mpc}}
\newcommand \kpc {h^{-1}{\rm kpc}}

\newcommand \farcm{\hbox{$.\!\!^{\prime}$}}
\newcommand \arcm{\hbox{$^{\prime}$}}

\newcommand \kms {{\rm km~s}^{-1}}

\newcommand \beqn {\begin{equation}}
\newcommand \eeqn {\end{equation}}

\newcommand \ROSAT {{\em ROSAT}~}
\documentclass[apj]{emulateapj}
\usepackage{epsfig}
\usepackage{natbib}

\begin{document}

\title{Infall Regions and Scaling Relations of X-ray Selected Groups}
\shorttitle{Infall Regions and Scaling Relations}
\shortauthors{Rines and Diaferio}

\author{Kenneth Rines\altaffilmark{1,2}
and Antonaldo Diaferio\altaffilmark{3,4}} 
\email{krines@cfa.harvard.edu}

\altaffiltext{1}{Smithsonian Astrophysical Observatory, 60 Garden St, MS 20, Cambridge, MA 02138; krines@cfa.harvard.edu}
\altaffiltext{2}{Department of Physics and Astronomy, Western Washington University, Bellingham, WA}
\altaffiltext{3}{Universit\`a degli Studi di Torino,
Dipartimento di Fisica Generale ``Amedeo Avogadro'', Torino, Italy; diaferio@ph.unito.it}
\altaffiltext{4}{Istituto Nazionale di Fisica Nucleare (INFN), Sezione di Torino, Torino, Italy}

\begin{abstract}

We use the Fifth Data Release of the Sloan Digital Sky Survey to study
X-ray-selected galaxy groups and compare their properties to clusters.
We search for infall patterns around the groups and use these to
measure group mass profiles to large radii.  In previous work, we
analyzed infall patterns for an X-ray-selected sample of 72 clusters
from the ROSAT All-Sky Survey.  Here, we extend this approach to a
sample of systems with smaller X-ray fluxes selected from the 400
Square Degree serendipitous survey of clusters and groups in ROSAT
pointed observations.  We identify 16 groups with SDSS DR5
spectroscopy, search for infall patterns, and compute mass profiles
out to 2-6 $\Mpc$ from the group centers with the caustic technique.
No other mass estimation methods are currently available at such large
radii for these low-mass groups, because the virial estimate requires
dynamical equilibrium and the gravitational lensing signal is too
weak.  Despite the small masses of these groups, most display
recognizable infall patterns.  We use caustic and virial mass
estimates to measure the scaling relations between different
observables, extending these relations to smaller fluxes and
luminosities than many previous surveys.  Close inspection reveals
that three of the groups are subclusters in the outskirts of larger
clusters.  A fourth group is apparently undergoing a group-group
merger.  These four merging groups represent the most extreme
outliers in the scaling relations.  Excluding these groups, we find
$L_X\propto\sigma_p^{3.4\pm1.6}$, consistent with previous
determinations for both clusters and groups.  Understanding cluster
and group scaling relations is crucial for measuring cosmological
parameters from clusters.  The complex environments of our group
sample reinforce the idea that great care must be taken in determining the
properties of low-mass clusters and groups.

\end{abstract}

\keywords{galaxies: clusters: individual  --- galaxies: 
kinematics and dynamics --- cosmology: observations }

\section{Introduction}

A large fraction of all galaxies are members of groups.  Groups are a
less extreme and much more common type of system than galaxy clusters.
Because they are less massive than clusters, groups are currently less
well understood than clusters.  Here, we investigate the optical
properties of a sample of X-ray-selected groups to determine whether
they can be modeled as scaled-down versions of clusters\footnote{The
division between groups and clusters is somewhat arbitrary; here, we
define groups to be systems of galaxies with virial masses smaller than
$10^{14} h^{-1} M_\odot$.}.  In particular, we study the outskirts of the
groups to determine if their infall regions are readily identifiable as
they are in the outskirts of clusters.  With large samples of
spectroscopic members, we then estimate the virial masses of the groups
and determine if the groups obey the same scaling relations as clusters.

Galaxy clusters and groups are surrounded by infall regions in which
galaxies are bound to the system but are not in equilibrium.  
The Cluster Infall Regions in the Sloan Digital Sky 
Survey \citep[][hereafter CIRS]{cirsi} project showed that 
X-ray-selected clusters display a characteristic trumpet-shaped pattern
in radius-redshift phase space diagrams.  These patterns, termed
caustics, were first predicted for simple spherical infall onto clusters
\citep{kais87,rg89}, but later work showed that these patterns reflect
the dynamics of the infall region \citep[][hereafter DG]{dg97} and
\citep[][hereafter D99]{diaferio1999}.
CIRS and earlier similar studies
\citep[e.g.,][]{gdk99,cairnsi} showed that the amplitude of
the caustics yields an estimate of cluster mass profiles consistent
with both virial and X-ray mass estimates where the techniques
overlap.  More recently, \citet{diaferio05} and \citet{lemze09} showed
that caustic masses also agree well with mass estimates based on
gravitational lensing.  Because neither galaxies nor gas is expected 
to be in equilibrium outside the virial radius of a cluster, the 
caustic technique and weak lensing are the only well-studied methods
for determining cluster mass profiles at large radii \citep[see][for 
a recent review of the caustic technique]{diaferio09}.

CIRS showed that infall patterns are ubiquitous in nearby massive 
clusters selected by X-ray emission.  
The CIRS clusters are fairly massive clusters and generally have
little surrounding large-scale structure \citep[but
see][]{rines01b,rines02}.  One might suspect that the presence of
infall patterns is limited to massive, isolated clusters.  However,
other investigators have found infall patterns around the Fornax
Cluster \citep{drink}, the Shapley Supercluster
\citep{rqcm}, an ensemble cluster comprised of poor clusters in the
Two Degree Field Galaxy Redshift Survey \citep{bg03}, and even the
galaxy group associated with NGC 5846 \citep{mahdavi05}. 
Here, we extend the study of infall patterns by studying a large
number of systems with smaller X-ray fluxes than the CIRS clusters.
Because these systems typically have smaller masses and fewer member
galaxies (more typical of groups), infall patterns might not be
identifiable in these systems.

In particular, we use the new 400 Square Degree (400d) survey of
clusters and groups in {\em ROSAT} PSPC pointed observations
\citep{burenin07}.  The 400d survey is the largest area cluster survey
extending to flux limits of $\approx$1.4$\times 10^{-13}$erg
s$^{-1}$cm$^{-2}$.  This flux limit is a factor of 20 smaller than the
flux limit of catalogs based on the {\em ROSAT} All-Sky Survey
\citep[RASS;][]{rass}.  We search for 400d clusters and groups in the
spectroscopic footprint of the Sloan Digital Sky Survey Data Release 5
\citep[][SDSS DR5]{dr5}.  This approach is similar to CIRS and to the
RASS-SDSS \citep{popesso04,popesso05} analysis of clusters in DR2.
Compared to both the CIRS and RASS-SDSS samples, the 400d groups are
expected to have significantly smaller masses on average.

One motivation for this study is to improve our understanding of
cluster scaling relations, in particular to extend these relations to
the regime of high-mass groups.  Ambitious cluster surveys like the
South Pole Telescope and Atacama Cosmology Telescope require a good
understanding of cluster scaling relations to accurately measure
cosmological parameters from the abundance and evolution of clusters.
Quantifying the scatter in these relations and any Malmquist-like bias
is crucial for obtaining robust cosmological constraints from cluster
surveys \citep[e.g.,][]{stanek06}.  Many forecasts for dark energy
constraints from cluster surveys assume that scatter in the
mass-observable scaling relations has a log-normal distribution
\citep[e.g.,][]{mantz08}.  Deviations from this assumption could
significantly impact the constraining power of these surveys.

Recent studies of scaling relations that extend into the group regime
have often reached contradictory conclusions.  For instance, a common
scaling relation is the $L_X-\sigma_p$ relation between the X-ray
luminosity $L_X$ of a cluster or group and its projected velocity
dispersion $\sigma_p$.  For massive clusters, this relation is
typically found to have a steep slope of 4.4$^{+0.7}_{-0.3}$
\citep{andilxsig} or 3.7$\pm$0.3 \citep{popesso05}.  For groups,
various studies have found slopes as shallow as 0.37 \citep[][ the 
fit is a broken power-law with a slope of 4.02 for clusters]{rasscals}
and as steep as 4.7$\pm$0.9 \citep{helsdon00b}.  A detailed
optical-X-ray study by \citet{osmond04} recently found a slope of
2.5$\pm$0.4.  The variety in slopes is produced by many factors,
including differing definitions of $L_X$, and in some cases, possible
evolution of the galaxies via dynamical friction \citep{helsdon05}.
One problem is that few large, complete group catalogs are available,
so many existing studies utilize either heterogeneous samples or
include many groups with limited data on each group.  One solution is
to utilize X-ray selection \citep[e.g.,][]{mulchaey98,rasscals}, but
existing cluster/group catalogs based on RASS
\citep[][CIRS]{popesso05} have been restricted to systems with
relatively high X-ray fluxes (and therefore including few groups).
The fainter X-ray flux limits of the 400d survey allow detection of
several groups.  Combining this catalog with optical data from the
Sloan Digital Sky Survey \citep{sdss}, we can test whether the scaling
relations show similar behavior for systems with smaller fluxes more
typical of galaxy groups.  A complementary approach (that we do not
apply here) to understanding scaling relations for low-mass systems is
to identify the systems optically and use stacking analysis to measure
their ensemble properties.  This approach has been applied previously 
with RASS data \citep{dai07,rykoff08}.

We describe the data and the group sample in $\S$ 2.  In $\S$ 3, we
review the caustic technique and use it to estimate the group mass
profiles, discuss cluster scaling relations, and compare the caustic
mass profiles to simple parametric models.  We discuss some individual
groups in $\S 4$. We compare the caustic mass profiles to X-ray and
virial mass estimators in $\S$ 5.  We discuss our results and conclude
in $\S 6$.  We assume $H_0 = 100~h~\kms \Mpc^{-1}, \Omega _m = 0.3,
\Omega _\Lambda = 0.7$ throughout.

\section{The 400d-SDSS Group Sample}

\subsection{Sloan Digital Sky Survey \label{sdssdesc}}

The Sloan Digital Sky Survey \citep[SDSS,][]{sdss} is a wide-area
photometric and spectroscopic survey at high Galactic latitudes.  The
Fifth Data Release (DR5) of SDSS includes 8000 square degrees of
imaging data and 5740 square degrees of spectroscopic data
\citep{dr5}.

The spectroscopic limit of the main galaxy sample of SDSS is $r$=17.77
after correcting for Galactic extinction \citep{strauss02}.  CAIRNS
and CIRS found that infall patterns were detectable in clusters
sampled to about $M^*+1$, or $z\lesssim$0.1 for SDSS data.  For the
current sample, the X-ray fluxes are much smaller than those of CAIRNS
or CIRS, so these groups may not be as well sampled.  If
the 400d groups are less massive on average, they should
contain fewer luminous galaxies than the CAIRNS and CIRS clusters.  We
therefore expect that infall patterns may be less common and/or poorly
sampled in the 400d groups.

Note that SDSS is $\sim$85-90\% complete to the nominal spectroscopic
limit.  The survey has $\approx$7\% incompleteness due to fiber
collisions \citep{strauss02}, which are more likely to occur in dense
cluster fields.  Because the target selected in a fiber collision is
determined randomly, this incompleteness can theoretically be
corrected for in later analysis. From a comparison of SDSS with the
Millennium Galaxy Catalogue, \citet{2004MNRAS.349..576C} conclude that there is an
additional incompleteness of $\sim$7\% due to galaxies misclassified
as stars or otherwise missed by the SDSS photometric pipeline.  For
our purposes, the incompleteness is not important provided sufficient
numbers of group galaxies do have spectra.

\subsection{X-ray Surveys for Clusters and Groups \label{xcs}}

In CIRS, we studied clusters contained in catalogs based on the ROSAT
All-Sky Survey \citep[RASS][]{rass}.  RASS is a shallow
survey but it is sufficiently deep to include nearby, massive
clusters and groups.  RASS covers virtually the entire sky and is 
thus the most complete X-ray cluster survey for nearby clusters and 
groups.  The flux limits of RASS-based surveys are $\approx$3$\times 10^{-12}$erg s$^{-1}$cm$^{-2}$ \citep{bcs,2000ApJS..129..435B}.

The wide field-of-view of the \ROSAT PSPC instrument provides a large
area for serendipitous discovery of sources in pointed observations.
Several cluster catalogs have been created with this goal.  The
largest of these catalogs is the 400d survey, which contains
essentially all pointed observations suitable for extragalactic
surveys \citep{burenin07}, yielding a total survey of about 400 square
degrees.  Extended X-ray sources are detected with a wavelet technique
and fluxes are measured in the ROSAT band (0.1-2.4 keV).  The flux
limit of the 400d survey is $\approx$1.4$\times 10^{-13}$erg
s$^{-1}$cm$^{-2}$, a factor of $\sim$20 fainter than the RASS-based
catalogs.  The effective area of the survey depends on the flux limit
and is calibrated with simulated observations of real ROSAT cluster
data.  The detection algorithm is optimized for luminous clusters
($L_X>5\times 10^{43}$erg s$^{-1}$ at moderate redshift ($z>0.3$).
The groups studied here have smaller X-ray luminosities and are thus
expected to be physically smaller than higher-luminosity clusters, but
they are also at much lower redshifts, so their angular sizes are not
too different from the luminous clusters at $z$$>$0.3.  The ROSAT
observations are typically not deep enough to separate diffuse group
emission from contributions from individidual galaxy halos.

We restrict our analysis to clusters and groups with $z\leq$0.10 ($\S
2.1$).  Our sample contains 16 groups within the SDSS DR5
spectroscopic footprint.  We will refer to this sample as the
400d-SDSS groups hereafter.  Figure \ref{cirslxz} shows the X-ray
luminosities of these systems compared to CIRS.  The range of $L_X$ in
the two surveys is similar, but the 400d-SDSS sample contains many
more systems with small $L_X$ than does CIRS.

\section{Results}

\subsection{Infall Patterns around X-ray Groups}

We first search for well-defined infall patterns around the 400d-SDSS
groups.  Analogously to CIRS, we plotted radius-redshift diagrams for
all groups in the 400d X-ray catalog covered by DR5 with
$z<$0.10.  We assign a ``by-eye'' classification of each group's
infall pattern: ``clean'' for groups with few background and
foreground galaxies, ``intermediate'' for groups with apparent
infall patterns but significant contamination from either related
large-scale structure or foreground and background objects, and
``weak'' for groups with little apparent infall pattern.  CIRS
successfully applied the caustic technique to clusters classified as
``intermediate'' or ``weak''; the classification scheme is thus fairly
conservative.  We use this classification scheme {\it only} to show
the dependence of the infall pattern appearance on group mass (using
$L_X$ as a proxy) and the sampling depth.

\begin{figure}
\figurenum{1}
\plotone{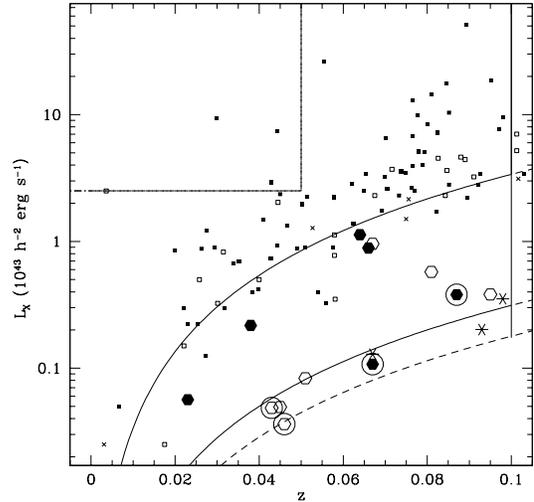}
\caption{\label{cirslxz} Redshift versus X-ray luminosity (0.1-2.4 keV) 
for X-ray clusters from CIRS (small squares) and 400d-SDSS groups
(large hexagons and stars) contained in the SDSS DR5 spectroscopic
survey region.  Filled hexagons, open hexagons and six-pointed stars
indicate 400d-SDSS groups with ``clean'', ``intermediate'', and
``weak'' infall patterns respectively.  These classifications for CIRS
clusters are represented by filled squares, open squares, and crosses.
Large open circles around some of the 400d-SDSS groups indicate
systems contaminated by nearby structure (see $\S \ref{individual}$).
The dashed line and lower solid lines show flux limits of $f_X>$1.4
and 3 $\times$10$^{-13}$erg s$^{-1}$ respectively; the effective
survey areas at these limits for the 400d survey are 232 and 370
square degrees respectively (not all of this area is contained within
the SDSS DR5 spectroscopic footprint).  The upper solid line shows the
flux and redshift limits of the CIRS cluster sample.  The dash-dotted
line shows the redshift and luminosity limits of CAIRNS. }
\end{figure}

Figure \ref{cirslxz} shows the dependence of this subjective
classification on X-ray luminosity and redshift.  The 400d-SDSS sample
contains 16 groups; six contain ``clean'' infall patterns and three
(19\%) show weak infall patterns.  The percentage in the last category
is larger than for the CIRS clusters (4\%), although the significance
of this difference is difficult to assess (the 400d-SDSS sample is
small and the categories are not robustly defined).

\begin{figure*}
\figurenum{2}
\plotone{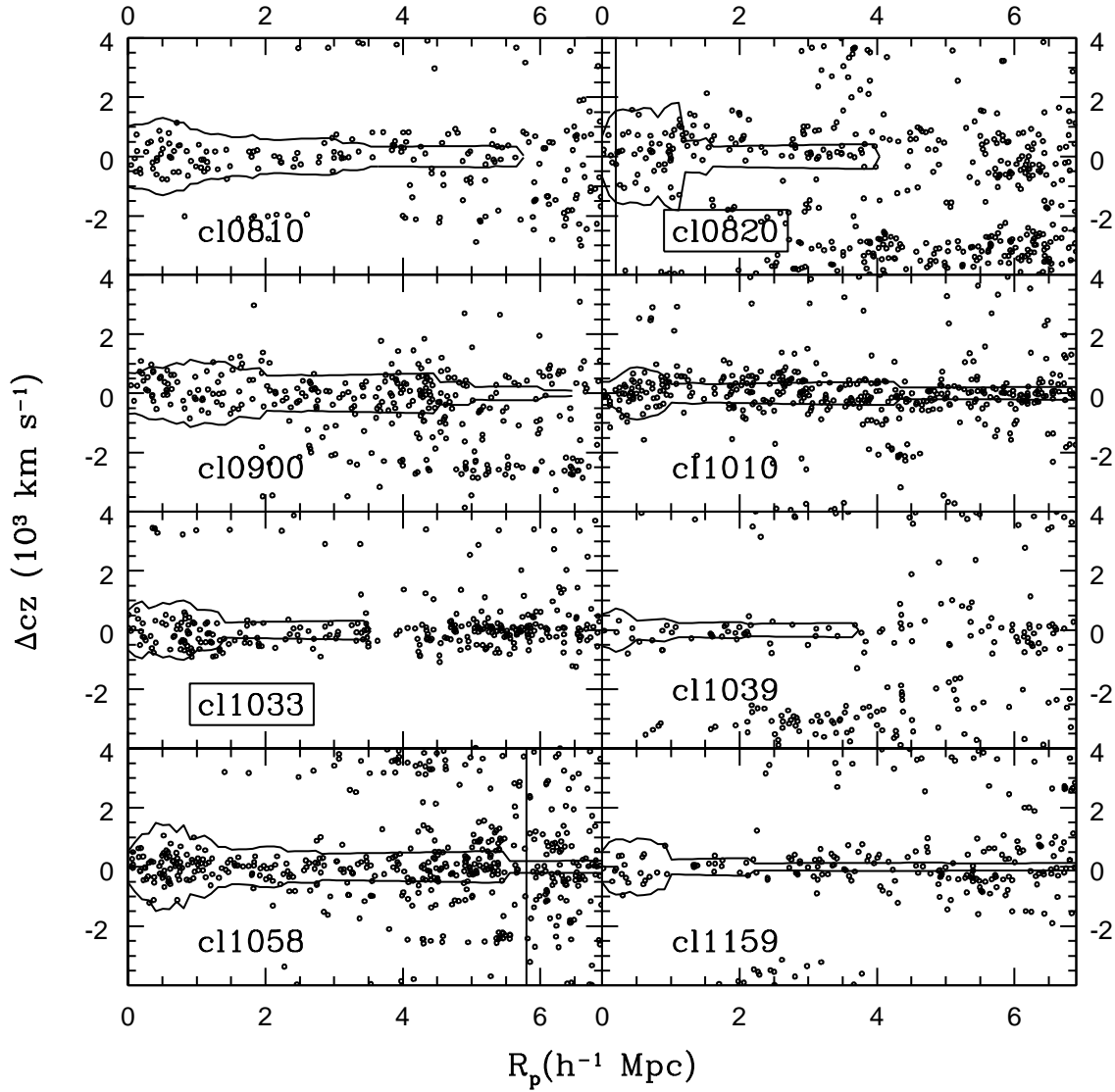}
\caption{\label{allcirs1} Redshift versus radius for SDSS galaxies around
the first eight X-ray groups in the 400d-SDSS sample.  The caustic
pattern is evident as the trumpet-shaped regions with high density.
The solid lines indicate our estimate of the location of the caustics
in each group.  
Vertical lines in each panel indicate the radius $R_{comp}$ where the
spatial coverage of the SDSS DR5 spectroscopic survey is no longer
complete (no line is shown if $R_{comp}>7\Mpc$).  Squares around
group names indicate systems with contamination from nearby groups
or clusters (see $\S \ref{individual}$).  Figure
\ref{allcirs2} shows similar plots for the rest of the sample.}
\end{figure*}

\begin{figure*}
\figurenum{3}
\plotone{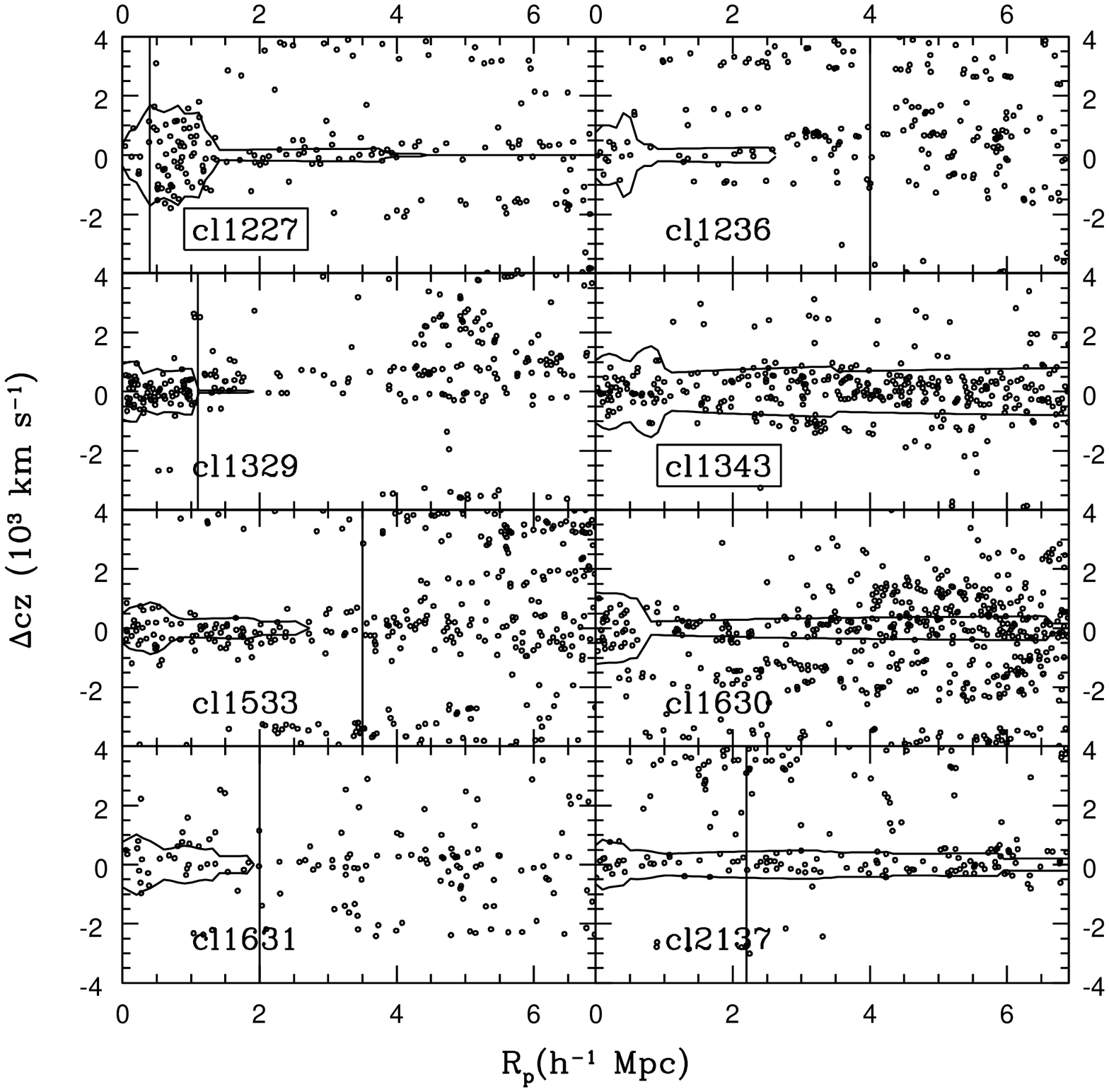}
\caption{\label{allcirs2} See Figure \ref{allcirs1}. }
\end{figure*}

Figure \ref{cirslxz} demonstrates the expanded parameter space covered
by the 400d-SDSS groups compared to the CIRS sample.  In particular, 
the 400d-SDSS sample includes many
more systems with $L_X$$\sim$10$^{42} h^{-2}$erg s$^{-1}$ than
CIRS.  Table \ref{sample} lists the groups in the 400d-SDSS
sample, their X-ray positions and luminosities, their central
redshifts and velocity dispersions (see below), and the projected
radius $R_{comp}$ within which the SDSS DR5 spectroscopic survey
provides complete spatial coverage.  For several groups, the caustic
pattern disappears beyond $R_{comp}$ because of this edge effect.
Figures \ref{allcirs1}-\ref{allcirs2} show the infall patterns for the
400d-SDSS sample.

\begin{table*}[th] \footnotesize
\begin{center}
\caption{\label{sample} \sc 400d-SDSS Basic Properties}
\begin{tabular}{llcccrrcccc}
\tableline
\tableline
\tablewidth{0pt}
400d & 400d & \multicolumn{2}{c}{X-ray Coordinates} & $z_\odot$ & $L_X /10^{43}$  & $\sigma_p$ & Flag & $R_{comp}$ \\ 
ID & Name & RA (J2000) & DEC (J2000) &   & erg~s$^{-1}$& $\kms$ & & $\Mpc$ \\ 
\tableline
         81 & cl0810+4216 & 122.60085 & 42.27195 & 0.0640 & 1.128 &  $ 473 ^{+73} _{-50}$  & 2 & $>$7  \\ 
         83\tablenotemark{a} & cl0820+5645 & 125.11083 & 56.75750 & 0.0430 & 0.049 &  $ 636 ^{+79} _{-57}$\tablenotemark{a}  & 1 & 0.2  \\ 
         89 & cl0900+3920 & 135.01958 & 39.34000 & 0.0950 & 0.382 &  $ 333 ^{+59} _{-39}$  & 1 & 0.7  \\ 
        106 & cl1010+5430 & 152.56584 & 54.50333 & 0.0462 & 0.052 &  $ 266 ^{+53} _{-33}$  & 1 & 2.1  \\ 
        112\tablenotemark{a} & cl1033+5703 & 158.46625 & 57.05306 & 0.0460 & 0.036 &  $ 397 ^{+81} _{-51}$\tablenotemark{a}  & 1 & 2.1  \\ 
        115 & cl1039+3947 & 159.88042 & 39.79389 & 0.0930 & 0.202 &  $ 193 ^{+132} _{-47}$  & 0 & 0.3  \\ 
        118 & cl1058+0136 & 164.55249 & 01.61583 & 0.0393 & 0.233 &  $ 415 ^{+44} _{-33}$  & 2 & 5.8  \\ 
        135 & cl1159+5531 & 179.96333 & 55.53222 & 0.0810 & 0.575 &  $ 321 ^{+79} _{-46}$  & 1 & 0.2  \\ 
        155\tablenotemark{a} & cl1227+0858 & 186.80875 & 08.97083 & 0.0901 & 0.409 &  $ 749 ^{+86} _{-64}$\tablenotemark{a}  & 2 & 0.4  \\ 
        159 & cl1236+1240 & 189.11917 & 12.40583 & 0.0440\tablenotemark{b} & 0.054 &  $ 384 ^{+117} _{-62}$  & 0 & 4.0  \\ 
        170 & cl1329+1143 & 202.36624 & 11.72306 & 0.0231 & 0.057 &  $ 363 ^{+46} _{-33}$  & 2 & 1.1  \\ 
        179\tablenotemark{a} & cl1343+5546 & 205.86624 & 55.78194 & 0.0685 & 0.112 &  $ 465 ^{+54} _{-40}$\tablenotemark{a}  & 2 & 0.3  \\ 
        199 & cl1533+3108 & 233.32126 & 31.14861 & 0.0670 & 0.958 &  $ 421 ^{+79} _{-51}$  & 1 & 3.5  \\ 
        204 & cl1630+2434 & 247.56125 & 24.58028 & 0.0637 & 0.823 &  $ 465 ^{+74} _{-50}$  & 2 & $>$7  \\ 
        205 & cl1631+2121 & 247.77042 & 21.36500 & 0.0952 & 0.331 &  $ 518 ^{+171} _{-89}$  & 0 & 2.0  \\ 
        227 & cl2137+0026 & 324.27875 & 00.44667 & 0.0506 & 0.082 &  $ 217 ^{+72} _{-38}$  & 1 & 2.2  \\ 
\tableline
\tablenotetext{a}{Strongly affected by a nearby group or cluster.  See $\S \ref{individual}$ for details.}
\tablenotetext{b}{\citet{burenin07} list the redshift as $z$=0.07.}
\end{tabular}
\end{center}
\tablecomments{Col.~(1) lists the number of the group in the 400d catalog 
\citep{burenin07}.  Col.~(2) lists the name of the group following the 
convention of \citet{burenin07}. Col.~(8) gives our subjective
classification of the contrast between the group and foreground and
background structure.  A Flag of 2 indicates a ``clean'' infall
pattern with little contaminating structure, while a Flag of 0
indicates a ``weak'' infall pattern for groups where the infall
pattern is weakly present and/or significantly contaminated by nearby
structure (see text for details).  Col.~(9) gives the projected radius
of the nearest edge of the footprint of the SDSS spectroscopic
survey.}
\end{table*}

Another factor which determines the presence or absence of a ``clean''
infall pattern is the surrounding large-scale structure.  For example,
the redshift-radius diagrams of groups within the infall regions of
more massive clusters reflect the kinematics of the cluster's infall
region
\citep[e.g., A2199, see][]{rines02}.  Closer inspection of the
400d-SDSS groups reveals that the caustic technique misidentifies
four of the 16 groups.  That is, the galaxy group associated with
the 400d X-ray source is embedded in a larger system containing more
galaxies.  The hierarchical clustering algorithm used by the caustic
technique to define the center of the system then mistakes the center
of the larger system as the center of the 400d group.  Ignoring these
four groups, the percentage of ``weak'' infall patterns increases
from 19\% to 25\% (three of the 12 remaining 400d-SDSS groups).
Although the statistics are poor, this result suggests that the
400d-SDSS groups might have less well-defined infall patterns than
the generally more massive CIRS clusters.

We attempt to quantify this difference more robustly by defining a
simple statistic to quantify the contrast of the infall pattern with
respect to the local background.  
We define the contrast
$C_{200}$ to be the ratio of the number of galaxies within projected
radius $r_{200}$ within the caustics $N_{mem}$ to the number of
``near-background'' galaxies $N_{bkgd}$, those within 
$\pm$5$\sigma_{200}$ of the cluster redshift but outside the caustics.
The radius $r_{200}$ is determined from the caustic
mass profile (see below) and $\sigma_{200}$ is the projected velocity
dispersion of galaxies within the caustics and within the projected
radius $r_{200}$.  Using this definition of contrast, there is
no significant correlation of $C_{200}$ with cluster mass for the CIRS
clusters.  Further, there is no significant
difference between the distributions of the $C_{200}$ statistic
between the 400d-SDSS (excluding the four problematic groups) and
CIRS samples (the samples differ at 59\% confidence using a K-S test).
This result suggests that any difference between the contrasts of the
two samples is too subtle for our simple contrast statistic.
We caution the reader that this contrast statistic is not well tested and
may have subtle dependencies on the caustic method used to define
membership.  It does provide a simple, quantitative way of comparing
the 400d-SDSS sample to the CIRS sample.

We discuss the individual groups in more detail in $\S \ref{individual}$.

\subsection{Caustics and Mass Profiles}


CAIRNS and CIRS showed that caustic masses of clusters agree well with
mass estimates from both X-ray observations and Jeans' analysis at
small radii, where dynamical equilibrium is expected to hold
\citep[][CIRS]{cairnsi}.  \citet{lokas03} confirm that the mass of
Coma estimated from higher moments of the velocity distribution agrees
well with the caustic mass estimate \citep{gdk99}.  Recently,
\citet{diaferio05} showed that caustic masses agree with weak
lensing masses in three clusters at moderate redshift.

We briefly review the method DG and D99 developed to estimate the mass
profile of a galaxy cluster by identifying caustics in redshift space.
The method assumes that clusters form in a hierarchical process.
Application of the method requires only galaxy redshifts and sky
coordinates.  Toy models of simple spherical infall onto clusters
produce sharp enhancements in the phase space density around the
system.  These enhancements, known as caustics, appear as a trumpet
shape in scatter plots of redshift versus projected clustercentric
radius \citep{kais87}.  DG and D99 show that random motions smooth out
the sharp pattern expected from simple spherical infall into a dense
envelope in the redshift-projected radius diagram \citep[see
also][]{vh98}.  The edges of this envelope can be interpreted as the
escape velocity as a function of radius.  Galaxies outside the
caustics are also outside the turnaround radius.  The caustic
technique provides a well-defined boundary between the infall region
and interlopers; one may think of the technique as a method for
defining membership that gives the cluster mass profile as a
byproduct.  In fact, the caustic technique can be used as a membership
classification for any gravitationally bound system.  For instance,
\citet{serra09} used caustics to identify interloper stars in five
dwarf spheroidal galaxies, and \citet{brown09} use caustics to
identify stars not bound to the Milky Way.

\begin{figure*}
\figurenum{4}
\epsscale{1.0}
\plotone{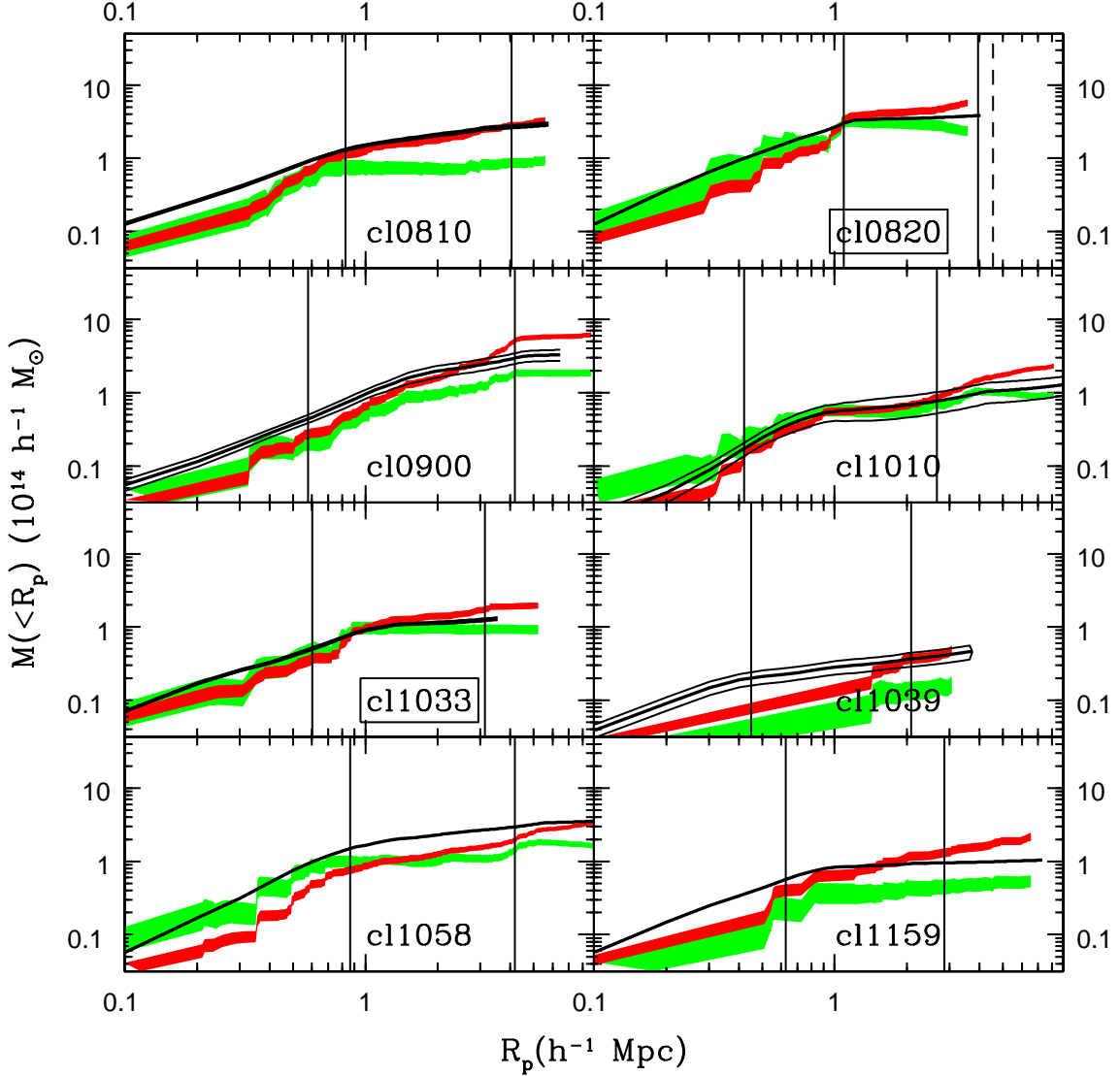}
\caption{\label{allcirsm1} Comparison of caustic mass profiles to those 
estimates from the virial theorem and the projected mass estimator.
The thick solid lines show the caustic mass profiles and the thin
lines show the 1$\sigma$ uncertainties in the mass profiles (for some
systems, the thin lines blend with the thick lines). The axes are
identical in all panels. The vertical bars indicate $r_{200}$ and the
maximum radius of the caustic mass profile (the smaller of $r_{max}$,
the extent of the infall pattern, and $r_t$, the turnaround radius).
Vertical dashed lines indicate $r_t$ for groups where the infall
pattern truncates before $r_t$. Red and green shaded regions show the
formal 1$\sigma$ uncertainties in the virial and projected mass
profiles.  Squares around group names indicate groups with
contamination from nearby groups or clusters (see $\S
\ref{individual}$).  Figure \ref{allcirsm2} shows similar plots for
the rest of the sample.}
\end{figure*}
\begin{figure*}
\figurenum{5}
\plotone{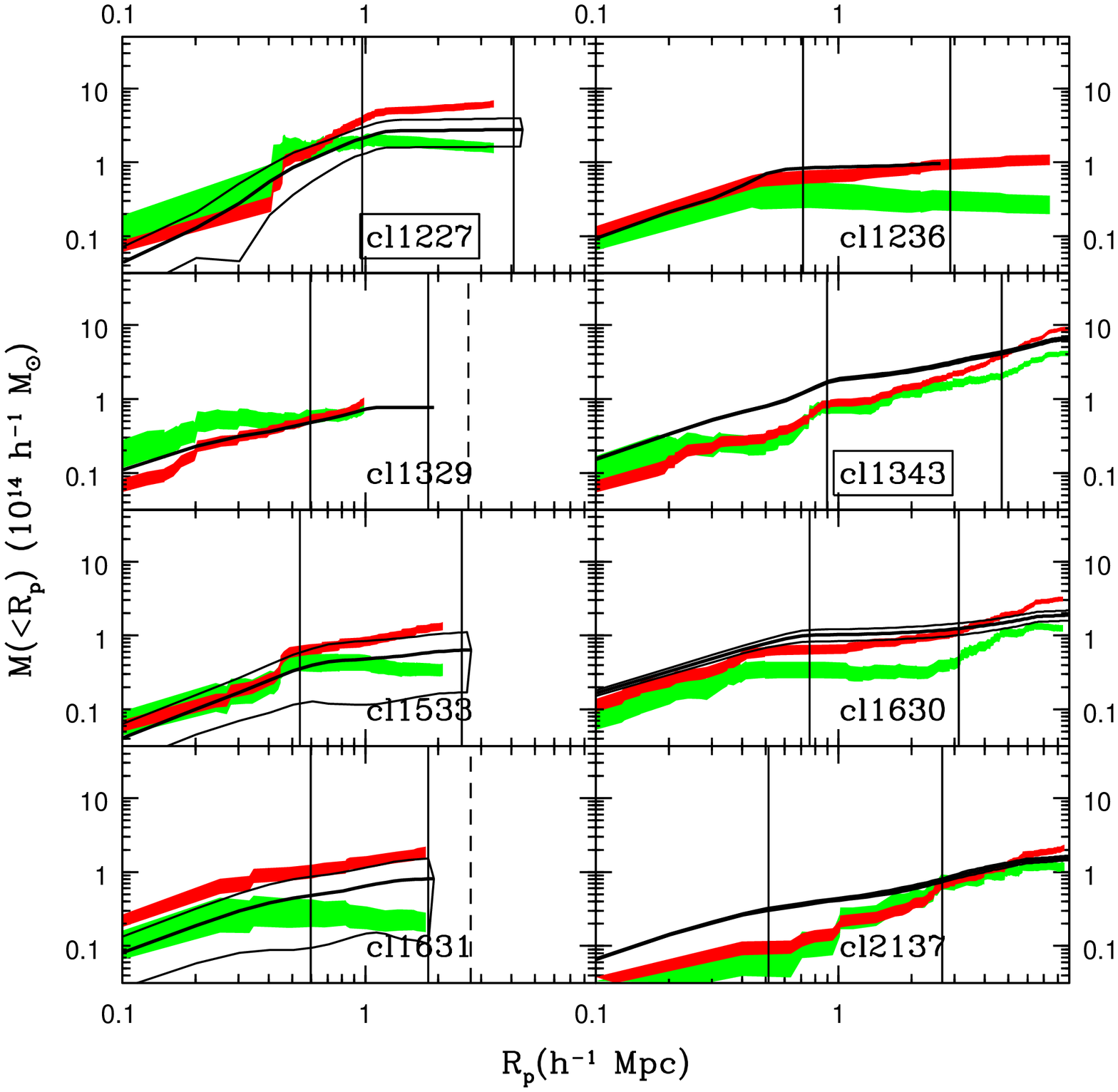}
\caption{\label{allcirsm2} See Figure \ref{allcirsm1}.}  
\end{figure*}

Operationally, we identify the caustics as curves which delineate a
significant decrease in the phase space density of galaxies in the
projected radius-redshift diagram.  
The details of the caustic technique used here are identical to those 
described in CIRS, with the following exception.  Because 400d-SDSS 
groups are expected to be less massive than CIRS clusters, we isolate 
the groups initially by studying only galaxies within
$R_p\leq$7$\Mpc$ and $\pm$4000$\kms$ of the nominal group centers
from the X-ray catalogs (compared to 10$\Mpc$ and $\pm$5000$\kms$ for 
the CIRS clusters).  We perform a hierarchical structure analysis
to locate the centroid of the largest system in each volume.  This
analysis sometimes finds the center of another system in the field.
In these cases, limiting the galaxies to a smaller radial and/or
redshift range enables the algorithm to center on the desired group.
For some groups, no satisfactory match between the hierarchical
center and X-ray center is possible; for these groups we impose the
X-ray center on the analysis (Table \ref{centers}).  
We discuss the groups individually in $\S$\ref{individual}.  Figures
\ref{allcirs1}-\ref{allcirs2} show the caustics and Figures
\ref{allcirsm1}-\ref{allcirsm2} show the associated mass profiles.
Note that the caustics extend to different radii for different
groups.  D99 shows that the appearance of the caustics depends
strongly on the line of sight; projection effects can therefore
account for most of the differences in profile shape in Figures
\ref{allcirs1}-\ref{allcirs2} without invoking non-homology among
clusters.  We use the caustics to determine group membership.  Here,
the term ``group member'' refers to galaxies both in the virial
region and in the infall region and inside the caustics.  Figures
\ref{allcirs1}-\ref{allcirs2} show that the caustics effectively
separate group members from background and foreground galaxies,
although some interlopers may lie within the caustics.  Note that, for
some groups, the small samples of galaxies yield unrealistically low
estimates of the uncertainties in the caustic mass profiles.

\begin{table*}[th] \footnotesize
\begin{center}
\caption{\label{centers} \sc 400d-SDSS Hierarchical Centers and Offsets From X-ray Centers}
\begin{tabular}{lcccr}
\tableline
\tableline
\tablewidth{0pt}
Group &\multicolumn{2}{c}{Hierarchical Center} & $\Delta cz$\tablenotemark{a} & $\Delta R$\tablenotemark{a}  \\ 
 & RA (J2000) & DEC (J2000) & $\kms$ & $\kpc$  \\ 
\tableline
         cl0810+4216 & 122.74532 & 42.33660 & 013 & 387   \\ 
         cl0820+5645\tablenotemark{b} & 125.11101 & 56.75749 & -- & --   \\ 
         cl0900+3920 & 135.02003 & 39.34003 & -- & --   \\ 
        cl1010+5430 & 152.55563 & 54.55338 & -366 & 115   \\ 
        cl1033+5703\tablenotemark{b} & 158.46603 & 57.05308 & -- & --   \\ 
        cl1039+3947 & 159.87998 & 39.79392 & -- & --   \\ 
        cl1058+0136 & 164.53830 & 01.62267 & -393 & 30   \\ 
        cl1159+5531 & 179.96301 & 55.53222 & -- & --   \\ 
        cl1227+0858\tablenotemark{b} & 186.87500 & 08.82286 & -34\tablenotemark{c} & --\tablenotemark{d}   \\ 
        cl1236+1240 & 189.06404 & 12.44854 & 002 & 149   \\ 
        cl1329+1143 & 202.32590 & 11.65740 & -17 & 89   \\ 
        cl1343+5546\tablenotemark{b} & 205.90356 & 55.62635 & -436 & 518   \\ 
        cl1533+3108 & 233.32102 & 31.14862 & -- & --   \\ 
        cl1630+2434 & 247.59420 & 24.57920 & 702 & 92   \\ 
        cl1631+2121 & 247.86281 & 21.51073 & 831 & --\tablenotemark{d}   \\ 
        cl2137+0026 & 324.25590 & 00.40342 & 127 & 121   \\ 
\tableline
\tablenotetext{a}{No value listed for the 6 groups where the X-ray center is used for the caustic analysis.}
\tablenotetext{b}{Strongly affected by a nearby group or cluster.  See $\S \ref{individual}$ for details.}
\tablenotetext{c}{\citet{burenin07} list a redshift of $z$=0.087 and reference \citet{burke03}.  However, \citet{burke03} list a redshift of $z$=0.090 for this group. }
\tablenotetext{d}{X-ray center adopted for spatial position and hierarchical central redshift.  The hierarchical centers are offset on the sky by $\Delta R$=685$~\kpc$ and 753$\kpc$ for cl1227+0858 and cl1631+2121 respectively.}
\end{tabular}
\end{center}
\end{table*}



The D99 algorithm we use to identify the caustics generally agrees
with the lines one would draw based on a visual impression.  This
consistency suggests that systematic uncertainties in the caustic
technique are dominated by projection effects rather than the details
of the algorithm.  We now discuss the individual groups in more detail.

\subsection{Comments on Individual Groups \label{individual}}

Groups share many common features, but each system is unique.  We
describe the most relevant aspects of each 400d-SDSS group below.

\begin{itemize}

\item {\it cl0810+4216}  The X-ray emission is centered on a $r$=14.2 
($M_r$$\approx$-22.3) galaxy.  The hierarchical center is located at
the position of a strongly bound galaxy pair 387$\kpc$ to the NE.  The
X-ray center corresponds to the position of the Brightest Group Galaxy
(BGG) and is probably the dynamical center of this system.  The infall
diagram produced when adopting the X-ray center as the system center
suggests that four sub-$L^*$ galaxies have slightly higher redshifts
than the bulk of the group members.  Excluding these galaxies from the
caustic-selected sample would reduce the inferred velocity dispersion
by $\sim$20\%.  The caustic technique classifies these four galaxies
as members using either the X-ray or the hierarchical center.

\item {\it cl0820+5645}  The X-ray emission is centered on a pair of 
interacting galaxies with $r$=15.7 and $r$=16.0.  A plot of group
members on the sky reveals two main clumps of galaxies, one on the
X-ray center and the other approximately centered on a $r=14.6$ galaxy (MCG
+09-14-024) without a redshift.  This system is possibly a group-group
merger.

\item {\it cl0900+3920}  The X-ray emission is centered $<$1$\arcm$ from 
a pair of galaxies with $r$=15.9 and $r$=17.1.  The Brightest Group
Galaxy ($r$=15.4) lies $\sim$4$\arcm$ north of the X-ray center.

\item {\it cl1010+5430}  The X-ray emission is centered on a $r$=14.2
galaxy, UGC 06057.  This galaxy has no SDSS redshift, but a redshift
from \citet{1995MNRAS.277.1312C} places it in the group.

\item {\it cl1033+5703}  The X-ray emission is centered on a $r$=14.2
galaxy.  The spatial distribution of galaxies near this group shows
that the X-ray group is located $\sim$800$\kpc$ from a group of
galaxies at the same redshift centered on CGCG 290-048 ($r$=13.5), a
radio galaxy with a jet 1.3$\Mpc$ in extent
\citep{1979MNRAS.187..253M}.  The 400d ROSAT pointing does not include
the CGCG 290-048 group, but it has been classified as an
optically-selected group by \citet{miller05} and \citet{merchan05}.
The dynamics of cl1033+5703 are probably perturbed by the proximity of
the CGCG 290-048 group \citep[for an analogous configuration, see
A2197E/W and A2199][]{rines02}.  Indeed, the CGCG 290-048 group
displays a more symmetric infall pattern than does cl1033+5703,
suggesting that the former system dominates the dynamics of the two
systems.

\item {\it cl1039+3947}  The X-ray emission is approximately centered 
on a $r$=15.0 galaxy.  A bright ($r$=15.2) spiral galaxy with several
fainter companions lies about 2$\Mpc$ NW of the group center.  This
latter system may be an infalling group.

\item {\it cl1058+0136}  The X-ray emission is centered on UGC 06057, 
a $r$=13.5 galaxy without a SDSS redshift but confirmed as a group
member using a redshift from \citet{rsmith00}.  This group is
6$\farcm$6 from the optical position of Abell 1139 \citep{aco1989} and
previously has been identified with A1139 by \citet{bcs} and
\citet{noras}.

\item {\it cl1159+5531}  The X-ray emission is centered on a $r$=14.1
($M_r$$\approx$-23.0) galaxy.  This group was classified as an X-ray
overluminous elliptical galaxy (OLEG) by \citet{1999ApJ...520L...1V}.

\item {\it cl1227+0858}  The X-ray emission is centered on a $r$=15.0
galaxy without a SDSS redshift.  A redshift from
\citet{1998ApJS..115....1S} shows this galaxy to be a member.  The
group lies $\sim$0.7$\Mpc$ from the Brightest Cluster Galaxy (BCG) of
Abell 1541, a cluster at the same redshift.  The hierarchical center
is coincident with the core of A1541, so we instead use the X-ray
center to estimate the caustics. \citet{noras} report an X-ray flux of
(2.5$\pm$0.5)$\times$10$^{-12}$erg s$^{-1}$cm$^{-2}$ for A1541,
$\sim$6 times larger than the flux from cl1227+0858.  This difference
suggests that A1541 dominates the dynamics of cl1227+0858 (see
cl1033+5703 above).  The dynamical parameters reported for cl1227+0858
thus partially reflect the properties of A1541, a more massive and
more X-ray luminous system.  Consistent with this idea, 7 galaxies
within 0.4$\Mpc$ of the X-ray center have a velocity dispersion
(594$^{+252}_{-111}\kms$) 20\% smaller than the value reported in
Table \ref{sample}; at slightly larger radii there is a large spike in
redshift space due to A1541 (Figure \ref{allcirs2}).


\item {\it cl1236+1240} This extended X-ray source is approximately 
centered on IC 3574, a $r$=14.1 ($M_r$$\approx$-22.4) galaxy at
$z$=0.067 \citep[][no SDSS spectrum available]{burenin07}.  SDSS
spectra, however, show that there are 15 galaxies at $z$$\approx$0.044
within 0.5$\Mpc$ of the X-ray center, while there are no other
galaxies within this radius at $z$$\approx$0.067.  The brightest of
these galaxies is $r$=15.5 ($M_r$$\approx$-20.2).  The X-ray emission
is therefore associated with a single bright galaxy (similar to a
fossil group, although most fossil groups do contain faint galaxies at
the same redshift) or with a group of less luminous galaxies (or some
combination).  We adopt the latter interpretation that cl1236+1240 is
associated with the group of less luminous galaxies.

\item {\it cl1329+1143}  The X-ray emission is centered between 
NGC 5179 ($r$=13.3) and NGC 5171 ($r$=12.9).  The latter has no SDSS
spectrum, but a redshift from \citet{falco99} places NGC 5171 in the
group.  This group is also known as MKW 11 and its infall region was
previously analyzed in CIRS.  Note that the X-ray flux from the 400d
catalog [(1.3$\pm$0.2)$\times$10$^{-12}$erg s$^{-1}$cm$^{-2}$] is
significantly smaller than the flux from RASS data
\citet[][]{noras} used in CIRS
[(5.7$\pm$0.8)$\times$10$^{-12}$erg s$^{-1}$cm$^{-2}$].  The smaller
flux from the 400d catalog would place this system below the flux
limit used in CIRS. \citet{osmond04b} analyze an XMM-Newton
observation of this group and determine a flux of
(2.5$\pm$0.4)$\times$10$^{-12}$erg s$^{-1}$cm$^{-2}$ in the (0.5-2.0
keV) band, intermediate between the RASS and 400d values.  A possible
explanation of the varying flux estimates is that a background group
southeast of NGC 5171 (centered on a red galaxy with SDSS photometric
redshift $z$=0.21$\pm$0.01) may have been included in the RASS flux,
while the 400d flux may exclude some emission associated with group
members. \citet{osmond04b} further find evidence for interactions
between the group members NGC 5171 and NGC 5176 and the surrounding
gas.  We find a slightly smaller velocity dispersion
(363$^{+46}_{-33}~\kms$) than do \citet[][494$\pm99~\kms$]{osmond04},
who noted that their value of $\sigma_p$ was larger than expected
based on the measured $T_X$=0.96$\pm$0.04 keV.  Our smaller value of
$\sigma_p$ mitigates this discrepancy.  The NGC 5129 X-ray group lies
$\sim$3$\Mpc$ northeast of cl1329+1143 at a similar redshift; it is
possible that these groups are physically associated.


\item {\it cl1343+5546}  The X-ray emission is centered on a $r$=15.8
galaxy.  The group is located 11$\farcm$4$\approx$500$\kpc$ from the
optical center of A1783.  The hierarchical center is coincident with
the core of A1783, so the reported dynamical properties reflect those
of the Abell cluster and not the 400d group.  Unlike cl1227+0858/A1541
above, cl1343+5546 and A1783 are too close together to obtain a mass
profile for cl1343+5546 alone.  \citet{2003AJ....126.2740L} report an
X-ray flux of (0.51$\pm$0.12)$\times$10$^{-12}$erg s$^{-1}$cm$^{-2}$
for A1783, $\sim$2.5 times larger than the flux of cl1343+5546.  The
galaxies in A1783 exhibit significant substructure, perhaps indicating
that the cluster is not yet relaxed.

\item {\it cl1533+3108}  The X-ray emission is centered on a $r$=15.3
galaxy possibly undergoing a merger with a $r$=16.7 galaxy; neither
galaxy has a SDSS spectrum.  A redshift from \citet{small97} places
the brighter galaxy in the group.  The central galaxy pair is
surrounded by several bright galaxies.  This group is also known as
Abell 2092.  \citet{wu99} list a velocity dispersion of 
$\sigma_p=504^{+115}_{-69}~\kms$, compared with our estimate of 
$\sigma_p=421^{+79}_{-51}~\kms$.  Our estimate is based on a larger 
number of galaxies. 

\item {\it cl1630+2434}  The X-ray emission is centered between a 
$r$=14.6 and a $r$=15.2 galaxy.  The brighter galaxy (IC 4607) has no
redshift available from either SDSS or the literature.

\item {\it cl1631+2121}  The X-ray center lies among seven red galaxies 
with $r$=16.5-18.0 (one a spectroscopically confirmed member); a
$r$=16.4 spiral galaxy (and spectroscopically confirmed member) lies
$\sim$1' northeast of the X-ray center.  The hierarchical center is
$\sim$750$\kpc$ northeast of the X-ray center, and $\sim$1$\farcm$5
southwest of a $r$=15.4 group member.  Many group members lie close to
the line connecting the X-ray and hierarchical centers.  We center the
mass profile on the X-ray center and the hierarchical redshift (Table
\ref{centers}).  Additional redshifts, especially of the galaxies near
the X-ray center, would help clarify the nature of this system.

\item {\it cl2137+0026}  The X-ray emission is centered between a 
$r$=14.3 early-type galaxy and a $r$=14.4 spiral galaxy.  Both are
group members, although the early-type galaxy has no SDSS spectrum
\citep[][]{falco99}.

\end{itemize}

\subsection{Virial and Turnaround Masses and Radii \label{virialm}}

The caustic mass profiles allow direct estimates of the virial and
turnaround radius in each group.  For the virial radius, we estimate
$r_{200}$ ($r_\Delta$ is the radius within which the enclosed average
mass density is $\Delta\rho _c$, where $\rho _c$ is the critical
density) by computing the enclosed density profile [$\rho (<r) = 3
M(<r)/4\pi r^3$]; $r_{200}$ is the radius which satisfies $\rho
(<r_{200})=200\rho_c$.  In our adopted cosmology, a system should be
virialized inside the slightly larger radius $\sim$$r_{100} \approx
1.3 r_{200}$ \citep{ecf96}.  We use $r_{200}$ because it is more
commonly used in the literature and thus allows easier comparison of
results.  For the turnaround radius $r_{t}$, we use equation (8) of
\citet{rg89} assuming $\Omega _m = 0.3$.  For this value of
$\Omega_m$, the enclosed density is 3.5$\rho _c$ at the turnaround
radius.  Varying $\Omega_m$ in the range 0.02--1 only changes the
inferred value of $r_t$ by $\pm$10\%; the uncertainties in $r_t$ from
the uncertainties in the mass profile are comparable or larger
\citep[D99;][]{rines02}.  If the $w$ parameter in the equation of
state of the dark energy ($P_\Lambda = w\rho_\Lambda$) satisfies $w\ge
-1$, the dark energy has little effect on the turnaround overdensity
\citep[][]{2002MNRAS.337.1417G}.

Table \ref{radii} lists $r_{200}$, $r_t$, and the masses $M_{200}$ and
$M_t$ enclosed within these radii.  For cl0820+0945, cl1329+1143, and
cl1631+21214, the maximum extent of the caustics $r_{max}$ is smaller
than $r_t$.  For these groups, $r_t$ is a minimum value assuming that
there is no additional mass outside $r_{max}$.  The best estimate of
the mass contained in infall regions clearly comes from those groups
for which $r_{max}\geq r_t$.  The average mass within the turnaround
radius for these groups is 2.4$\pm$0.4 times the virial mass
$M_{200}$, demonstrating that groups are still forming in the present
epoch and consistent with the CIRS estimate of 2.19$\pm$0.18.
Similarly, the average turnaround radius is (5.01$\pm$0.25)$r_{200}$
for groups with $r_{max}\geq r_t$, again in agreement with the CIRS
result of (4.96$\pm$0.08)$r_{200}$.  Simulations of the future growth
of large-scale structure \citep{2002MNRAS.337.1417G,nl02,busha03} for
our assumed cosmology ($\Omega _m = 0.3, \Omega _\Lambda = 0.7$)
suggest that galaxies currently inside the turnaround radius of a
system will continue to be bound to that system.  Our results for the
turnaround radius and mass agree with the predictions of
\citet{busha05}, who find that the ultimate mass of dark matter haloes
in simulations is 1.9$M_{200}$.  The agreement between the estimates
of $M_t/M_{200}$ and $r_t/r_{200}$ between the CIRS and 400d-SDSS
samples suggests that the overall shapes of cluster mass profiles into
their infall regions are not strongly dependent on cluster mass into 
the group regime.  We
compare the virial masses $M_{200}$ from the caustics with masses
calculated using the virial theorem in $\S \ref{virialsec}$.

\begin{table*}[th] \footnotesize
\begin{center}
\caption{\label{radii} \sc 400d-SDSS Characteristic Radii and Masses}
\begin{tabular}{lccccccr}
\tableline
\tableline
\tablewidth{0pt}
Group & $r_{500}$ & $r_{200}$ & $r_t$ & $r_{max}$ & $M_{200}$ & $M_{t}$ & $M_{t}/M_{200}$ \\ 
 & $\Mpc$ & $\Mpc$ & $\Mpc$ & $\Mpc$ & $10^{14} M_\odot$ & $10^{14} M_\odot$ &  \\ 
\tableline
         cl0810+4216 & 0.52 & 0.83 & 4.05 & 9.49 & 1.32$\pm$ 0.05 & 2.70$\pm$0.14 & 2.05   \\
         cl0820+5645\tablenotemark{a} & 0.67 & 1.09 & $>$4.55 & 3.94 & 3.03$\pm$ 0.02 & 3.84$\pm$0.02 & 1.27   \\
         cl0900+3920 & 0.35 & 0.58 & 4.17 & 9.19 & 0.45$\pm$ 0.06 & 2.95$\pm$0.48 & 6.58   \\
        cl1010+5430 & 0.20 & 0.42 & 2.66 & 9.80 & 0.17$\pm$ 0.04 & 0.77$\pm$0.25 & 4.45   \\
        cl1033+5703\tablenotemark{a} & 0.38 & 0.60 & 3.14 & 9.60 & 0.51$\pm$ 0.02 & 1.26$\pm$0.06 & 2.48   \\
        cl1039+3947 & 0.29 & 0.45 & 2.08 & 3.64 & 0.21$\pm$ 0.04 & 0.37$\pm$0.08 & 1.74   \\
        cl1058+0136 & 0.51 & 0.86 & 4.17 & $>$10 & 1.51$\pm$ 0.00 & 2.96$\pm$0.00 & 1.96   \\
        cl1159+5531 & 0.39 & 0.63 & 2.86 & 9.29 & 0.57$\pm$ 0.00 & 0.95$\pm$0.00 & 1.67   \\
        cl1227+0858\tablenotemark{a} & 0.56 & 0.97 & 4.09 & 4.34 & 2.13$\pm$ 0.83 & 2.79$\pm$1.16 & 1.31   \\
        cl1236+1240 & 0.49 & 0.71 & 2.88 & 9.70 & 0.84$\pm$ 0.00 & 0.97$\pm$0.00 & 1.16   \\
        cl1329+1143 & 0.40 & 0.59 & $>$2.66 & 1.82 & 0.48$\pm$ 0.00 & 0.77$\pm$0.00 & 1.58   \\
        cl1343+5546\tablenotemark{a} & 0.53 & 0.90 & 4.70 & $>$10 & 1.69$\pm$ 0.06 & 4.23$\pm$0.20 & 2.50   \\
        cl1533+3108 & 0.32 & 0.54 & 2.50 & 2.63 & 0.36$\pm$ 0.23 & 0.64$\pm$0.47 & 1.76   \\
        cl1630+2434 & 0.52 & 0.76 & 3.12 & 9.09 & 1.03$\pm$ 0.18 & 1.24$\pm$0.23 & 1.20   \\
        cl1631+2121 & 0.40 & 0.60 & $>$2.72 & 1.82 & 0.49$\pm$ 0.39 & 0.82$\pm$0.71 & 1.67   \\
        cl2137+0026 & 0.35 & 0.51 & 2.67 & 8.99 & 0.32$\pm$ 0.01 & 0.78$\pm$0.05 & 2.45   \\
\tableline
\tablenotetext{a}{Strongly affected by a nearby group or cluster.  See $\S \ref{individual}$ for details.}
\end{tabular}
\end{center}
\tablecomments{Cols.~(1)-(3) list $r_\delta$ for three values of 
overdensity $\delta$, where $r_\delta$ is the radius within which the
enclosed mass density is $\delta$ times the critical density (we adopt
$\delta$=3.5 as the turnaround overdensity).  Col.~(4) gives the
turnaround radius $r_t$ for each group.  Col.~(5) gives the maximum
radius $r_{max}$ within which the caustics are detected.  Note that
when $r_{max}<r_t$, the estimate of $r_t$ is an underestimate if any
mass lies between $r_{max}$ and $r_t$.}
\end{table*}

One striking result of this analysis is that the caustic pattern is
often visible beyond the turnaround radius of a group (similar to some 
CIRS clusters).  This result
suggests that even groups may have strong dynamic effects on surrounding
large-scale structure beyond the turnaround radius.  For our assumed
cosmology, this large-scale structure is probably not bound to the
group.  

\subsection{Comparison to Virial and Projected  Mass Estimates \label{virialsec}}


We apply the virial mass and projected mass estimators
\citep{htb} to the 400d-SDSS groups.  For the latter, we assume the
galaxies are on isotropic orbits.  We must define a radius of
virialization within which the galaxies are relaxed.  We use $r_{200}$
(Table \ref{radii}) and include only galaxies within the caustics.  We
thus assume that the caustics provide a good division between group
galaxies and interlopers (see Figures \ref{allcirs1}-\ref{allcirs2}).  
For details on these mass calculations, see CIRS.
Table \ref{virial} lists the virial and projected mass estimates. 

\begin{table*}[th] \footnotesize
\begin{center}
\caption{\label{virial} \sc 400d-SDSS Virial and Projected Masses}
\begin{tabular}{lcrrr}
\tableline
\tableline
\tablewidth{0pt}
Group & $r_{200}$ & $M_{200}$ & $M_{proj}$ & $M_{vir}$  \\ 
 & $\Mpc$  & $10^{14} M_\odot$ & $10^{14} M_\odot$   & $10^{14} M_\odot$  \\ 
\tableline
         cl0810+4216 & 0.83 & 1.32$\pm$ 0.05 & 0.79$\pm$0.19 & 1.14$\pm$0.17   \\
         cl0820+5645\tablenotemark{a} & 1.09 & 3.03$\pm$ 0.02 & 3.15$\pm$0.65 & 3.15$\pm$0.41   \\
         cl0900+3920 & 0.58 & 0.45$\pm$ 0.06 & 0.21$\pm$0.06 & 0.27$\pm$0.04   \\
        cl1010+5430 & 0.42 & 0.17$\pm$ 0.04 & 0.20$\pm$0.06 & 0.12$\pm$0.02   \\
        cl1033+5703\tablenotemark{a} & 0.60 & 0.51$\pm$ 0.02 & 0.45$\pm$0.14 & 0.35$\pm$0.06   \\
        cl1039+3947 & 0.45 & 0.21$\pm$ 0.04 & 0.08$\pm$0.04 & 0.18$\pm$0.00   \\
        cl1058+0136 & 0.86 & 1.51$\pm$ 0.00 & 1.02$\pm$0.18 & 0.76$\pm$0.09   \\
        cl1159+5531 & 0.63 & 0.57$\pm$ 0.00 & 0.24$\pm$0.08 & 0.41$\pm$0.08   \\
        cl1227+0858\tablenotemark{a} & 0.97 & 2.13$\pm$ 0.83 & 1.09$\pm$0.21 & 3.58$\pm$0.45   \\
        cl1236+1240 & 0.71 & 0.84$\pm$ 0.00 & 0.39$\pm$0.16 & 0.59$\pm$0.12   \\
        cl1329+1143 & 0.59 & 0.48$\pm$ 0.00 & 0.58$\pm$0.12 & 0.52$\pm$0.07   \\
        cl1343+5546\tablenotemark{a} & 0.90 & 1.69$\pm$ 0.06 & 0.79$\pm$0.15 & 0.85$\pm$0.11   \\
        cl1533+3108 & 0.54 & 0.36$\pm$ 0.23 & 0.46$\pm$0.13 & 0.63$\pm$0.10   \\
        cl1630+2434 & 0.76 & 1.03$\pm$ 0.18 & 0.35$\pm$0.09 & 0.64$\pm$0.10   \\
        cl1631+2121 & 0.60 & 0.49$\pm$ 0.39 & 0.31$\pm$0.13 & 0.97$\pm$0.20   \\
        cl2137+0026 & 0.51 & 0.32$\pm$ 0.01 & 0.07$\pm$0.03 & 0.09$\pm$0.02   \\
\tableline
\tablenotetext{a}{Strongly affected by a nearby group or cluster.  See $\S \ref{individual}$ for details.}
\end{tabular}
\end{center}
\tablecomments{Cols.~(4) and (5) give the projected and virial mass estimates for each group computed with all galaxies inside the radius $r_{200}$ determined from the caustic mass profile.}
\end{table*}

\begin{figure}
\figurenum{6}
\plotone{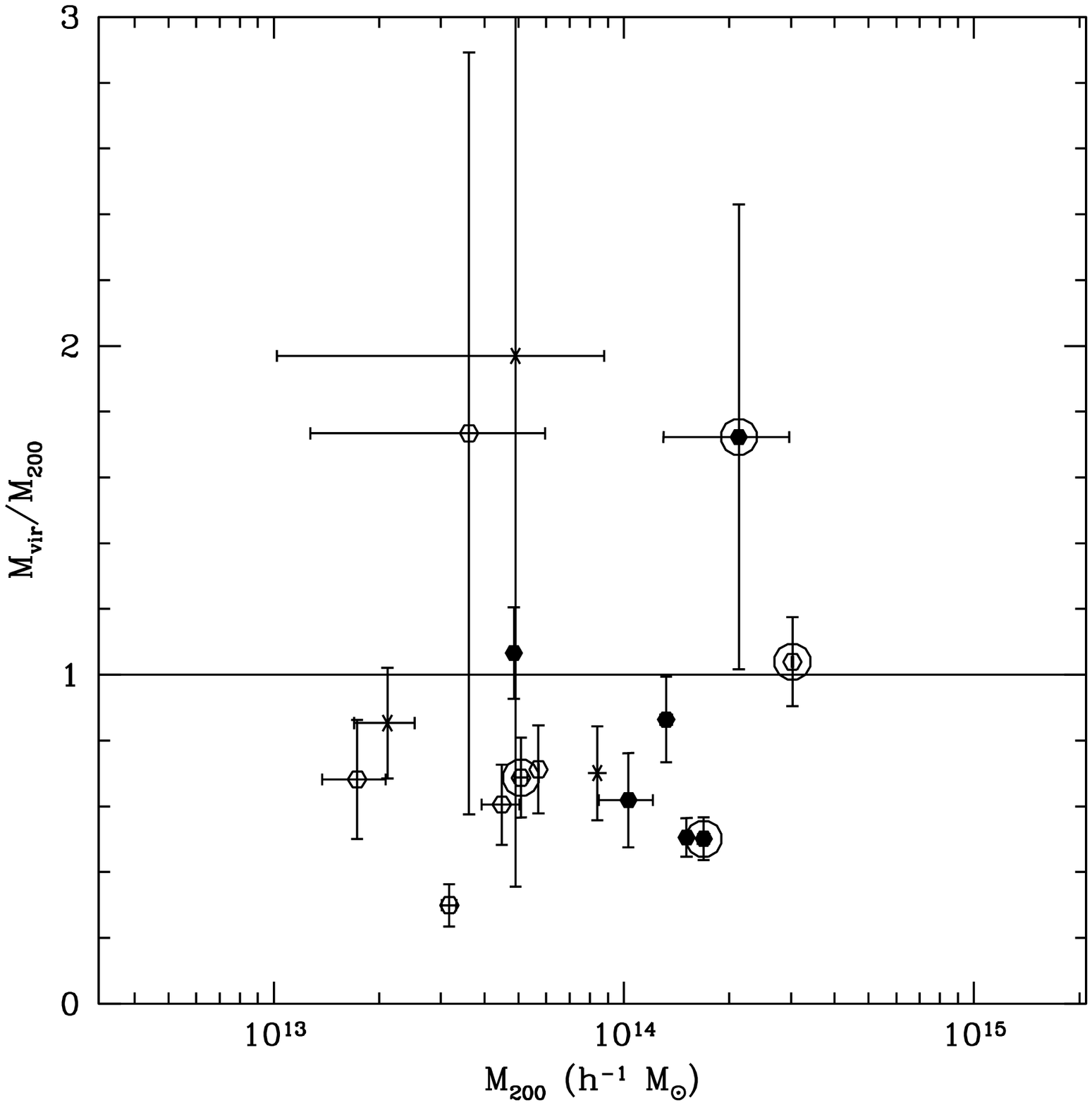} 
\caption{\label{cirsvc} Caustic masses at $r_{200}$ compared to 
virial masses at the same radius.  Symbol types are the same as for
Figure 1.  Four groups that are strongly affected by nearby systems
are highlighted with large circles.  Errorbars show 1$\sigma$
uncertainties and the solid line shows a ratio of unity.
}
\end{figure}

Figure \ref{cirsvc} compares the virial and caustic mass estimates at
$r_{200}$.  The mean ratios of these estimates are $M_v/M_c = 0.91 \pm
0.12$. 
The caustic mass estimates are
consistent with virial mass estimates even assuming a correction
factor $C\approx 0.1-0.2 M_{vir}$ for the surface pressure, consistent
with the best-fit NFW profiles \citep[see
also][CIRS]{cye97,girardi98,kg2000,cairnsi}.  Note that
\citet{lemze09} suggest a possible modification of the form factor
$\mathcal{F}_\beta$ used in the caustic technique based on a
comparison of the caustic and lensing mass profiles of A1689.
\citet{lopes09} find a smaller scatter in scaling relations using
virial masses rather than caustic masses for the CIRS clusters.

Figures \ref{allcirsm1}-\ref{allcirsm2} compare the mass profiles
estimated from the caustics, virial theorem, and projected mass
estimator.  The virial and projected mass profiles $M(<R_P)$ are 
calculated using all galaxies within the projected radius
$R_P$ (note that the virial assumptions hold only at the virial
radius). The projected mass estimator consistently overestimates the
mass at small radii and underestimates the mass at large radii
relative to the other profiles.  This behavior suggests that this
estimator is best for estimating virial masses but not mass profiles.
The virial and caustic mass profiles generally agree although there
are many groups with large disagreements.  The caustic mass profiles
do not appear to consistently overestimate or underestimate the mass
relative to the virial mass profiles.  This result supports our use of
caustic mass profiles as a tracer of the total group mass profile
in the next section.

\subsection{The Shapes of Group Mass Profiles \label{shapes}}

We fit the mass profiles of the 400d-SDSS groups to three simple
analytic models.  The simplest model of a self-gravitating system is a
singular isothermal sphere (SIS). The mass of the SIS increases
linearly with radius.  \citet[][]{nfw97} and
\citet{hernquist1990} propose two-parameter models based on CDM
simulations of haloes.  We note that the caustic mass profiles mostly
sample large radii and are therefore not very sensitive to the inner
slope of the mass profile.  Thus, we do not consider alternative
models which differ only in the inner slope of the density profile
\citep[e.g.,][]{moore99}.  At large radii, the best constraints on
cluster mass profiles come from galaxy dynamics and weak lensing.  The
caustic mass profiles of Coma
\citep{gdk99}, A576 \citep{rines2000}, A2199 \citep{rines02} and the
rest of the CAIRNS and CIRS clusters \citep{cairnsi,cirsi} provided
strong evidence against a singular isothermal sphere (SIS) profile and
in favor of steeper mass density profiles predicted by \citet[][NFW]{nfw97} and \citet{hernquist1990}.  Only recently have weak
lensing mass estimates been able to distinguish between SIS and NFW
density profiles at large radii \citep{clowe01,kneib03,mandelbaum06}.

At large radii, the NFW mass profile increases as ln$(r)$ and the mass
of the Hernquist model converges.  The NFW mass profile is
\beqn
M(<r) = \frac{M(a)}{\mbox{ln}(2) - \frac{1}{2}}[\mbox{ln}(1+\frac{r}{a})-\frac{r}{a+r}]
\eeqn
where $a$ is the scale radius and $M(a)$ is the mass within $a$. We
fit the parameter $M(a)$ rather than the characteristic density
$\delta_c$ (${M(a) = 4\pi \delta_c \rho_c a^3 [\mbox{ln}(2) -
\frac{1}{2}]}$ where $\rho_c$ is the critical density) because $M(a)$
and $a$ are much less correlated than $\delta_c$ and $a$
\citep{mahdavi99}.  The Hernquist mass profile is
\beqn
M(<r) = M \frac{r^2}{(r+a_H)^2}
\eeqn
where $a_H$ is the scale radius and $M$ is the total mass. Note that
$M(a_H) = M/4$. The SIS mass profile is $M(<r)\propto r$.
We minimize $\chi ^2$ and list the best-fit parameters $a$, $r_{200}$,
the concentration $c_{200}$=$r_{200}/a$, and $M_{200}$ for the
best-fit NFW model and indicate the best-fit profile type in Table
\ref{mpfitsci}.  We also list the parameter $c_{101}$=$r_{101}/a$;
some authors prefer to use $r_{101}$ as the virial radius for this
cosmology.  We perform the fits on all data points within the maximum
radial extent of the caustics $r_{max}$ listed in Table \ref{radii}
and with caustic amplitude $\mathcal{A}
\mathnormal{(r)} > 100~\kms$.

\begin{table*}[th] \footnotesize
\begin{center}
\caption{\label{mpfitsci} \sc 400d-SDSS Mass Profile Fit Parameters}
\begin{tabular}{lcccccc}
\tableline
\tableline
\tablewidth{0pt}
Group & $a_{NFW}$ & $r_{200}$ & $c_{200}$ & $M_{200}$ & Best-fit & $c_{101}$\\ 
 & $\Mpc$ & $\Mpc$ & & $10^{14} M_\odot$ &  Profile \\ 
\tableline
       cl0810+4216 & 0.185 & 0.79 & 4.26 & 1.14 & N & 5.75   \\ 
       cl0820+5645\tablenotemark{a} & 0.205 & 0.95 & 4.63 & 1.98 & H & 6.23   \\ 
       cl0900+3920 & 0.590 & 0.60 & 1.02 & 0.50 & N & 1.49   \\ 
      cl1010+5430 & 0.415 & 0.40 & 0.96 & 0.15 & H & 1.42   \\ 
      cl1033+5703\tablenotemark{a} & 0.160 & 0.61 & 3.83 & 0.53 & H & 5.19   \\ 
      cl1039+3947 & 0.072 & 0.41 & 5.75 & 0.16 & H & 7.68   \\ 
      cl1058+0136 & 0.297 & 0.77 & 2.59 & 1.06 & H & 3.58   \\ 
      cl1159+5531 & 0.160 & 0.58 & 3.64 & 0.46 & H & 4.95   \\ 
      cl1227+0858\tablenotemark{a} & 0.415 & 0.75 & 1.80 & 1.01 & H & 2.59   \\ 
      cl1236+1240 & 0.081 & 0.62 & 7.70 & 0.57 & H & 10.20   \\ 
      cl1329+1143 & 0.064 & 0.60 & 9.32 & 0.49 & N & 12.30   \\ 
      cl1343+5546\tablenotemark{a} & 0.200 & 0.86 & 4.29 & 1.46 & N & 5.78   \\ 
      cl1533+3108 & 0.138 & 0.48 & 3.46 & 0.25 & H & 4.71   \\ 
      cl1630+2434 & 0.051 & 0.68 & 13.40 & 0.75 & H & 17.56   \\ 
      cl1631+2121 & 0.092 & 0.58 & 6.32 & 0.47 & H & 8.52   \\ 
      cl2137+0026 & 0.081 & 0.50 & 6.20 & 0.30 & N & 8.27   \\ 
\tableline
\tablenotetext{a}{Strongly affected by a nearby group or cluster.  See $\S \ref{individual}$ for details.}
\end{tabular}
\end{center}
\tablecomments{Cols.~(2) and (3) give the scale radius $a_{NFW}$ and $r_{200}$ of the best-fit NFW profile.  Cols.~(4) and (6) give the NFW concentration parameters $c_\delta = r_\delta/a_{NFW}$ where $r_\delta$ is the radius within which the enclosed density is $\delta$ times the critical density.}
\end{table*}

Because the individual points in the mass profile are not independent,
the absolute values of $\chi ^2$ are indicative only, but it is clear
that the NFW and Hernquist profiles provide acceptable fits to the
caustic mass profiles; the SIS is never the best-fit profile and it is
actually excluded for nearly all groups.  The NFW profile provides a
better fit to the data than the Hernquist profile for 5 of the 16
400d-SDSS groups; the remaining 11 are better fit by a Hernquist
profile.

\begin{figure*}
\figurenum{7}
\plottwo{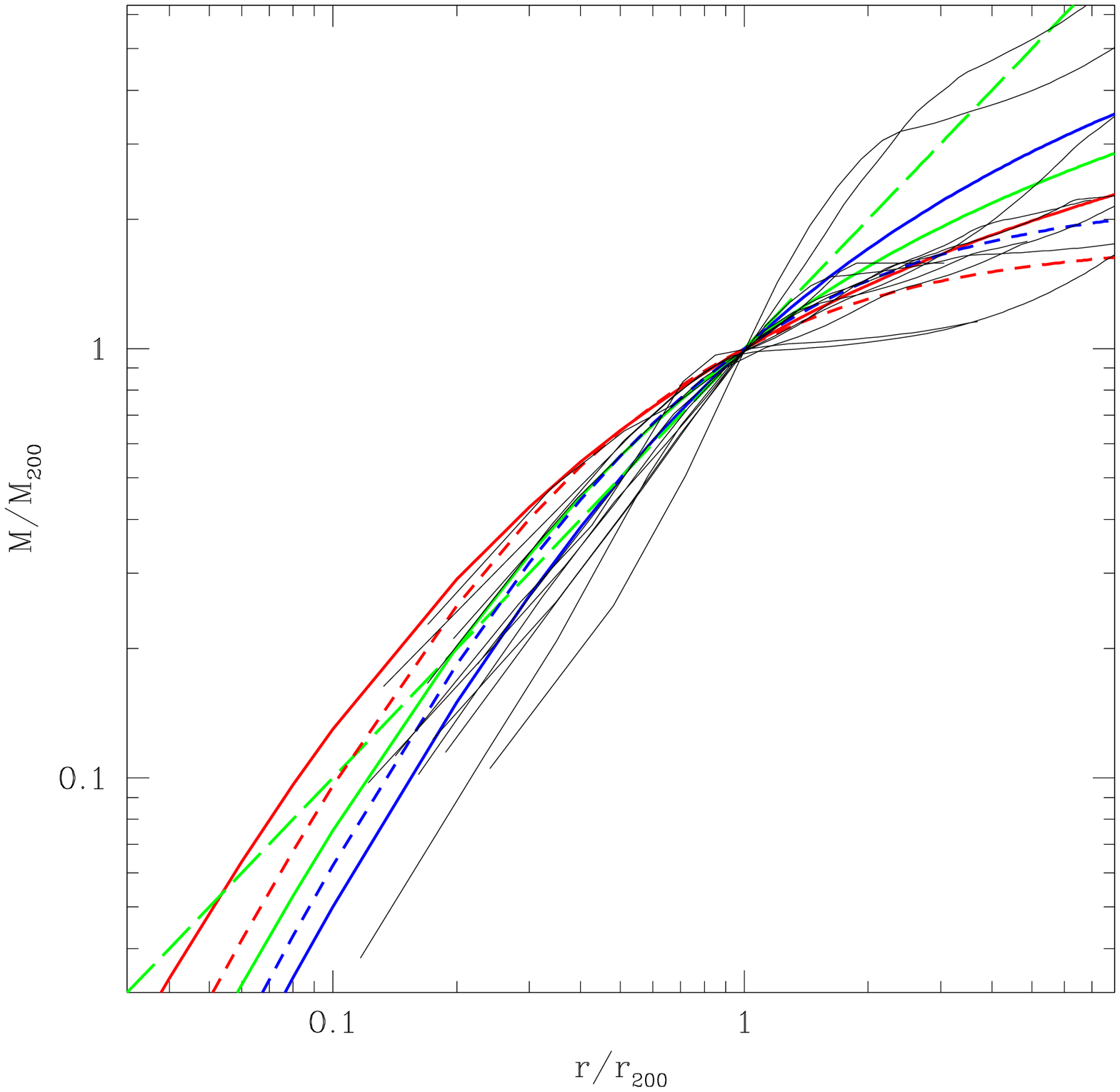}{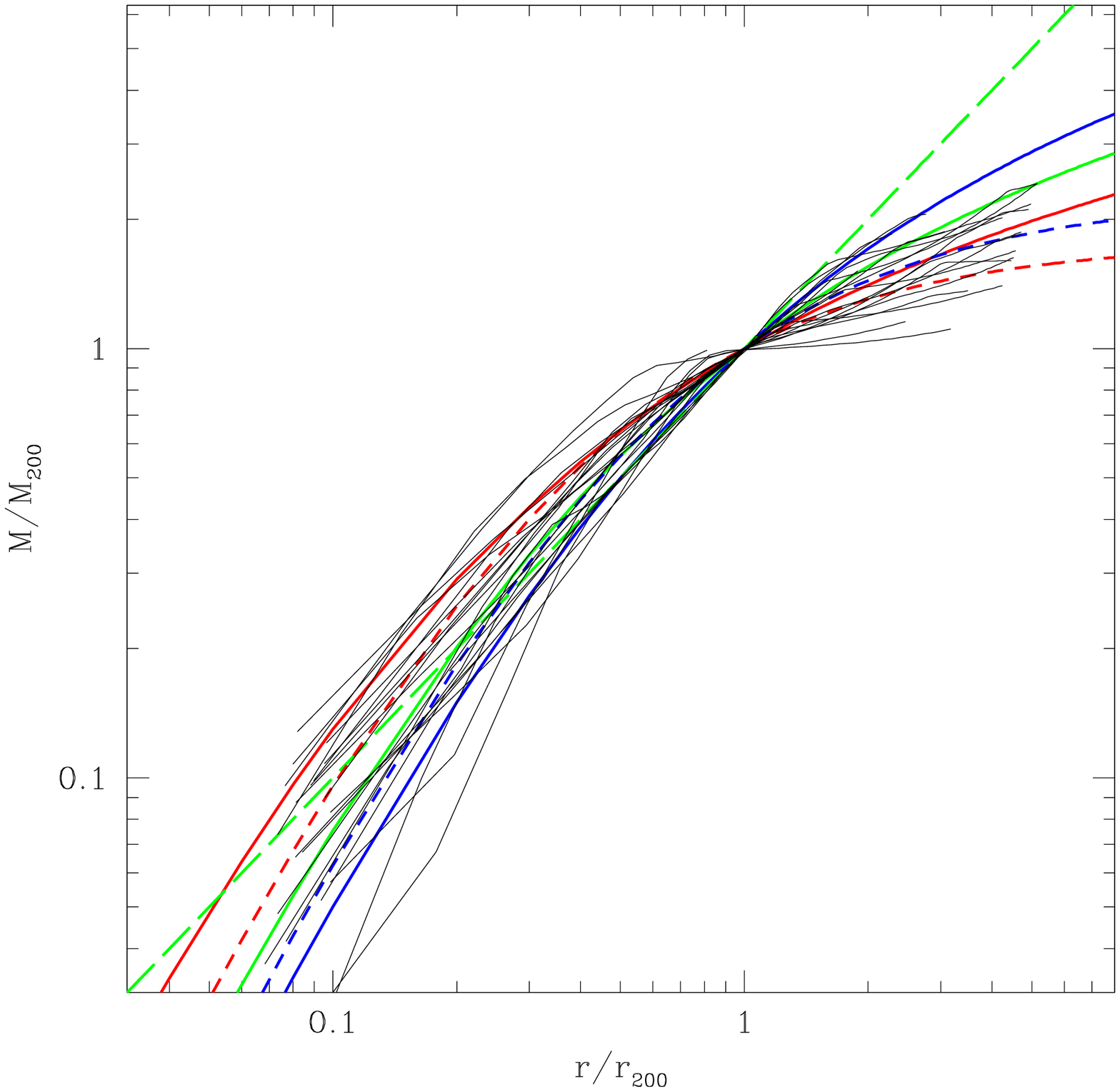}
\caption{\label{scalem} Scaled caustic mass profiles for the 400d-SDSS
groups (excluding those contaminated by nearby large-scale structure) 
compared to simple models. The thin solid lines show the
caustic mass profiles normalized by $r_{200}$ and $M_{200}$.  The
long-dashed line shows a singular isothermal sphere, the colored solid
lines show NFW profiles (with concentrations $c_{200}$=3,5,10 from top to
bottom at large radii).  The short-dashed lines are Hernquist profiles
with scale radii different by a factor of two.  Right: Scaled caustic 
mass profiles for the most massive quartile (ordered by $M_{200}$) of 
CIRS clusters with the same model profiles.}
\end{figure*}

Figure \ref{scalem} shows the shapes of the caustic mass profiles
scaled by $r_{200}$ and $M_{200}$ along with SIS, NFW, and Hernquist
model profiles.  The colored lines show different model mass profiles.
The straight dashed line is the SIS, the solid lines are NFW profiles
with $c_{200}$=3,5, and 10, and the curved dashed lines are Hernquist
profiles with two different scale radii.  The average value of
$c_{200}$ is 6.5, consistent with CIRS and with the values expected
from simulations for clusters \citep[NFW;][]{bullock01}.  All
three model profiles agree fairly well with the caustic mass profiles
in the range (0.1-1)$r_{200}$.  The SIS only fails beyond
$\sim$1.5$r_{200}$; this is why lensing has had trouble distinguishing
between SIS and NFW profiles.  As discussed in D99, the caustic
technique can be subject to large variations for individual groups
due to projection effects.  The best constraints on the shapes of
cluster and group mass profiles are obtained by averaging over many lines of
sight, or for real observations, over many different clusters and groups.

\begin{figure}
\figurenum{8}
\plotone{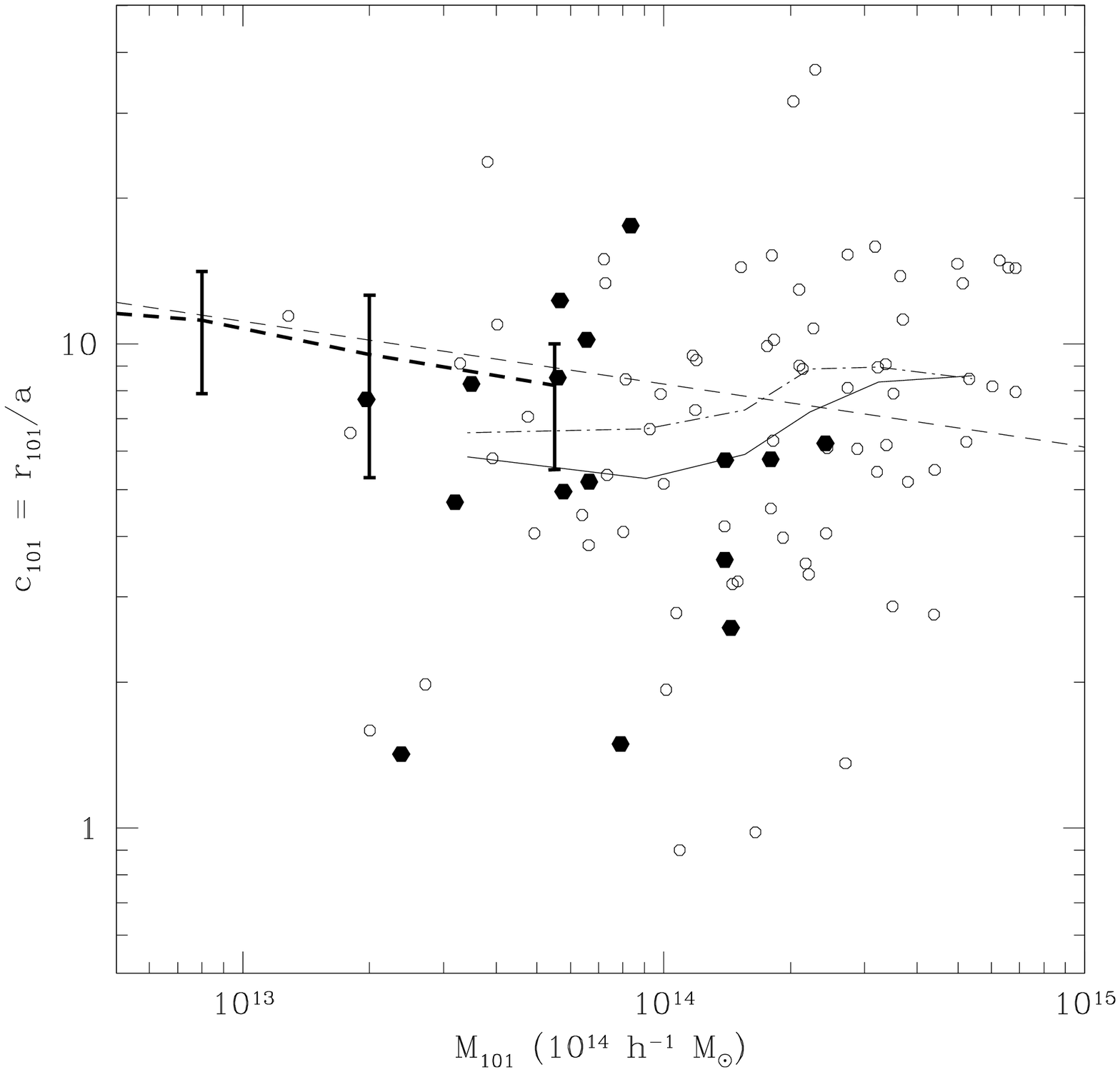} 
\caption{\label{cnfw} Concentrations $c_{101}$=$r_{101}/a$ of the best-fit NFW 
profiles verus $M_{101}$ for the 400d-SDSS groups (solid hexagons)
and the CIRS clusters (open circles).  The dashed lines show the
expected trend from (thick dashed line) simulations and a parametric model \citep[thin dashed line][]{bullock01}. The solid and
dash-dotted lines show the median value of $c_{101}$ and
$\mbox{log}(c_{101})$ for the CIRS clusters as a function of cluster
mass. }
\end{figure}

The concentration parameters $c_{200}=r_{200}/a$ for the NFW models
are in the range 1-15 (Table \ref{mpfitsci}), in good agreement with
the predictions of numerical simulations \citep{nfw97}.  Numerical
simulations suggest that $c$ should decrease slightly ($\sim$30\%)
with increasing mass over this mass range, but this decrease is 
comparable to the scatter in $c$ in simulated clusters
\citep{nfw97,bullock01}.  The errorbars in Figure \ref{cnfw} indicate
the average values and $1\sigma$ scatter in $c_{101}=r_{101}/a$ in
simulations \citep{bullock01}.  

Observations of mass profiles using X-ray mass estimates
\citep{buote07} and weak lensing \citep{comerford07} appear to confirm
a decrease in $c$ with increasing mass, although some massive clusters
apparently have high concentrations \citep[e.g.][]{lemze09}.  In
contrast, the CIRS clusters showed a weak positive correlation of $c$
with mass (Figure \ref{cnfw}).  The 400d-SDSS groups do not confirm
this correlation (Figure \ref{cnfw}).  A Spearman rank-sum test
indicates that there is no significant correlation between $c_{101}$
and $M_{101}$ (20\% confidence) for the 400d-SDSS groups.  A Spearman
test for the CIRS clusters suggests a correlation at the 92.7\%
confidence level.  By combining the CIRS and 400d-SDSS data, a
Spearman test indicates a correlation between concentration and mass
at the 93.6\% confidence level (correlation coefficient 0.19).  This
result indicates possible tension between caustic mass profiles and
simulations, as the caustic mass profiles indicate that concentrations
increase rather than decrease with increasing cluster mass.  More data
for groups at lower masses would help clarify the situation.

To further explore the dependence of mass profile shapes on mass, the
right panel of Figure \ref{scalem} shows the scaled mass profiles of the
most massive quartile of CIRS clusters.  Comparing the two panels
reveals no obvious trend with increasing mass: the profiles of 400d-SDSS
groups have larger scatter, but the average of the profiles appears
similar to that of the massive CIRS clusters.

\section{Scaling Relations of Clusters and Groups} 

Scaling relations between simple cluster/group observables and masses provide
insight into the nature of cluster/group assembly and the properties of
various cluster/group components.  Establishing these relations for local
clusters and groups is critical for future studies of clusters in the distant
universe with the goal of constraining dark energy
\citep{majumdar04,lin04}.  

We apply the prescription of \citet{danese} to determine the mean
redshift $cz_\odot$ and projected velocity dispersion $\sigma_p$ of
each group from all galaxies within the caustics.  We calculate
$\sigma_p$ using only the group members projected within $r_{200}$
estimated from the caustic mass profile.  Note that our estimates of
$r_{200}$ do not depend directly on $\sigma_p$.

One of the simplest X-ray observables of clusters and groups is X-ray
luminosity.  For optical data, the projected velocity dispersion
$\sigma_p$ is a straightforward observable.  In particular,
\citet{lopes09} reanalyze the CIRS clusters and find that the velocity
dispersions are robust to the particular choice of interloper removal
algorithms, while their virial masses have significant scatter
relative to the CIRS values (primarily due to differences in estimates
of $r_{200}$).  The $L_X -\sigma_p$ relation is expected to have the
form $L_X \propto \sigma_p^4$ under simple theoretical models
\citep{quintana82}.  For clusters, most recent estimates of the slope
of this relation are consistent with a slope of 4.  For galaxy groups,
however, some authors find shallower slopes \citep{rasscals,xue00},
while others find slopes comparable to clusters
\citep{ponman96,mulchaey98}.  In the Group Evolution Multiwavelength
Survey (GEMS), \citet{osmond04} find a slope of 2.31$\pm$0.61 for
groups, but the slope for a combined sample of groups and clusters is
4.55$\pm$0.25. In a similar vein, \citet{rasscals} find that the
best-fit relation for a sample of groups and clusters is a broken
power-law with a shallower slope for groups than for clusters.
Unusual systems may affect the $L_X -\sigma_p$ relation: for instance,
\citet{helsdon05} show that galaxy groups with apparently very small
$\sigma_p$ and large $L_X$ may have larger $\sigma_p$ than earlier
estimates, bringing these groups more in line with a steeper
relation.  \citet{khosroshahi07} find that fossil groups follow a
$L_X-\sigma_p$ relation with slope 2.74$\pm$0.45, shallower than
clusters (they also note a tendency for fossil groups to have slightly
larger $L_X$ for a given $\sigma_p$ than non-fossil groups).  In the
400d-SDSS sample, we selected the groups based on their X-ray
emission, whereas most previous studies
\citep{ponman96,mulchaey98,rasscals,osmond04} begin with catalogs of
optically selected groups and then analyze either pointed X-ray
observations or RASS data.

\begin{figure}
\figurenum{9}
\plotone{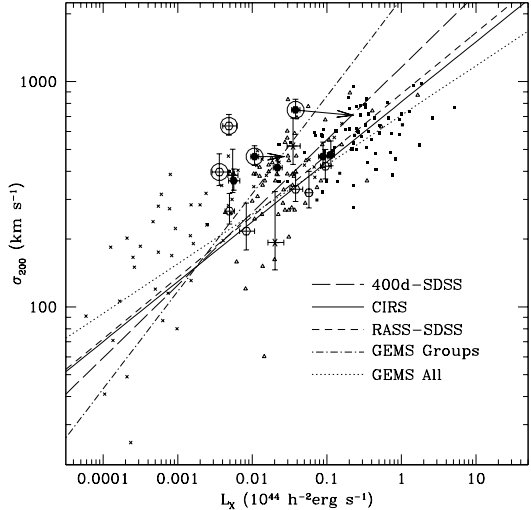} 
\caption{\label{lxsigma} Velocity dispersions at $r_{200}$ compared to X-ray 
luminosities. The solid line is the bisector of the least squares fits
for the CIRS clusters.  The dashed lines show the $\sigma_{p}-L_X$
relations from RASS-SDSS \citep{popesso05}.  Symbol types are
explained in Figure 1; large open circles highlight groups strongly
affected by nearby structures.  The two arrows indicate the locations 
of these data points for the larger clusters located close to two of 
these groups (A1541 near cl1227+0858 and A1783 near cl1343+5546).}
\end{figure}
   
The 400d-SDSS X-ray luminosities are measured in the ROSAT band
(0.1-2.4 keV) and corrected for Galactic absorption \citep[see][for
details]{burenin07}.  Figure \ref{lxsigma} shows the $L_X -\sigma_p$
relation for the CIRS clusters along with the best-fit $L_X -\sigma_p$
relation of the RASS-SDSS \citep{popesso05} and CIRS clusters, the
GEMS groups, and the GEMS groups+clusters fit \citep{osmond04}.  The
400d-SDSS groups roughly follow the relations of the RASS-SDSS and
CIRS samples, but the scatter is large.  Squares, triangles, and
crosses in Figure \ref{lxsigma} show systems from CIRS,
\citet{rasscals}, and \citet{osmond04} respectively.  Note that the
400d observations are typically not deep enough to separate group
emission from contributions from individual galaxy halos, so the X-ray
luminosities are not necessarily directly comparable to those in some
other studies \citep[e.g.,][]{rasscals,osmond04}.  However, because
the 400d-SDSS groups lie close to RASSCALS and GEMS groups, this
effect is probably small.


The groups previously identified as undergoing mergers are
highlighted with large circles in Figure \ref{lxsigma}.  These groups
are the most extreme outliers from the CIRS and RASS-SDSS scaling
relations and the locus of points from earlier studies.  This result
can be easily understood as a result of confusing an $L_X$ measured
for a subcluster with the velocity dispersion $\sigma_p$ reflecting
the velocity dispersion of the parent cluster.  Consistent with this
explanation, the arrows in Figure \ref{lxsigma} show where cl1227+0858
(cl1343+5546) would appear if we used the X-ray luminosity of the
nearby, more luminous system A1541 (A1783); both points move closer to
the CIRS $L_X-\sigma$ relation.  Note that only one of the
400d-SDSS groups (cl1159+5531) is a likely fossil group
\citep{vikhlinin98}, so the sample is not dominated by these unusual
systems.  Specifically, \citet{voevodkin09} search for fossil groups
and clusters in the 400d survey; cl1159+5531 is the only fossil group at
$z$$<$0.1, and it only qualifies under a relaxed definition of
fossil groups.

Excluding the four groups contaminated by large-scale structure, 
the bisector of the
least-squares fits has a slope of 3.4$\pm$1.6, intermediate between
the shallow slopes found by some investigators for groups and the
steeper slopes found for clusters (see Figure \ref{lxsigma}).  Because
the uncertainty in the slope for the 400d-SDSS groups is large, the
slope is not significantly different from any of these relations.

We test the consistency of the $L_X-\sigma_p$ relation for 400d-SDSS
groups with that of the CIRS clusters by fitting these relations for
the CIRS clusters alone and for the CIRS clusters with the 400d-SDSS
groups (excluding those contaminated by large-scale structure) added
to the sample.  For the CIRS data alone, the slope of the bisector of
least-squares fits is $3.7\pm0.6$, and the slope for the combined CIRS
and 400d-SDSS data is $4.1\pm0.4$.

From the limited sample, we conclude that the X-ray-selected 400d-SDSS
groups have an $L_X-\sigma_p$ relation consistent with previous
studies of both larger-mass clusters (e.g., CIRS) and smaller-mass
groups (e.g., GEMS).  We find no evidence of large biases in this
relation between X-ray-selected and optically-selected systems.  An
important caveat to this conclusion is that some recent studies of
clusters \citep{popesso06} and groups \citep{rasmussen06} find that
there may be some systems that are significantly ``X-ray
underluminous,'' perhaps because these systems are dynamically young.


\begin{figure*}
\figurenum{10}
\plottwo{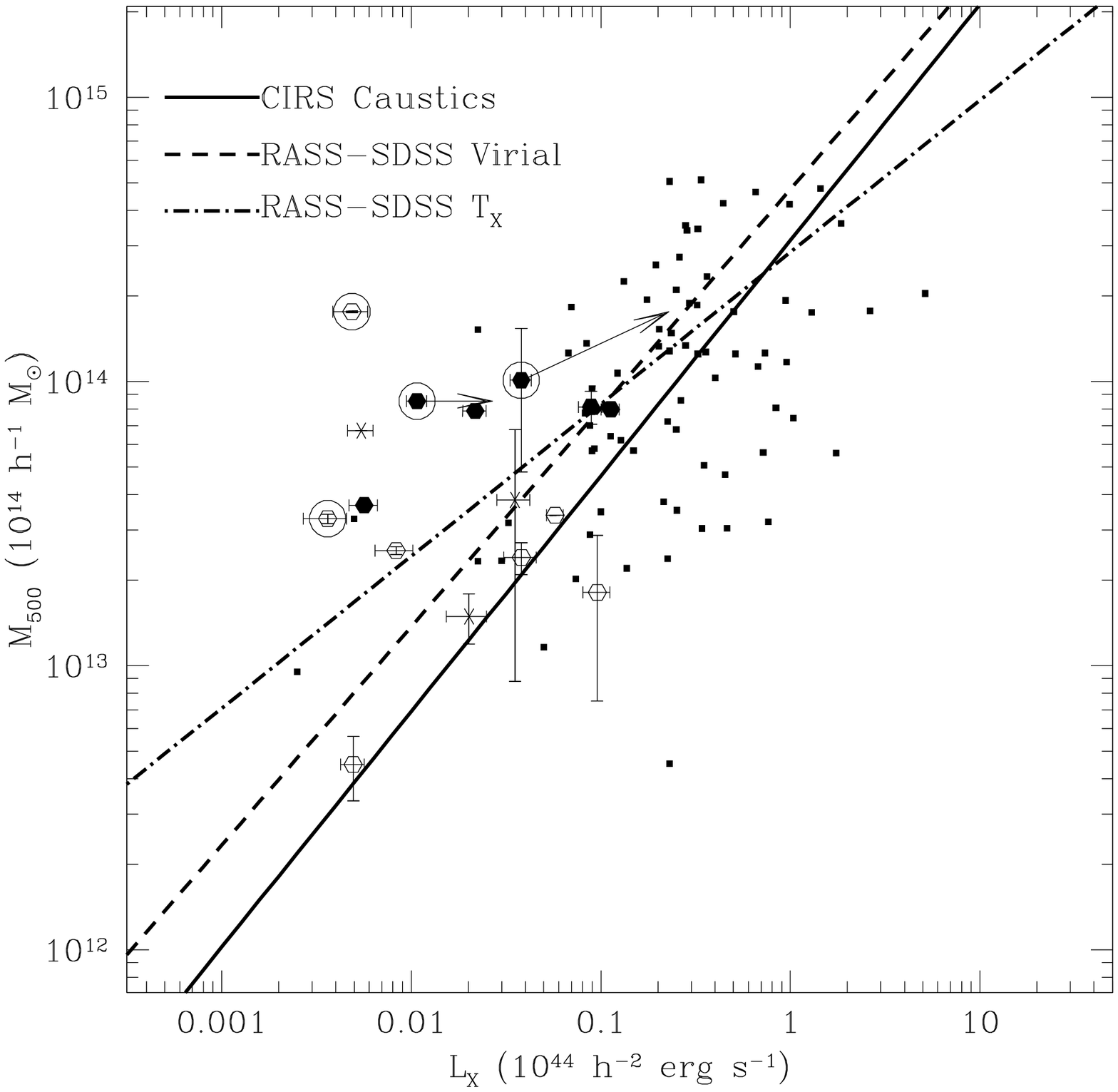}{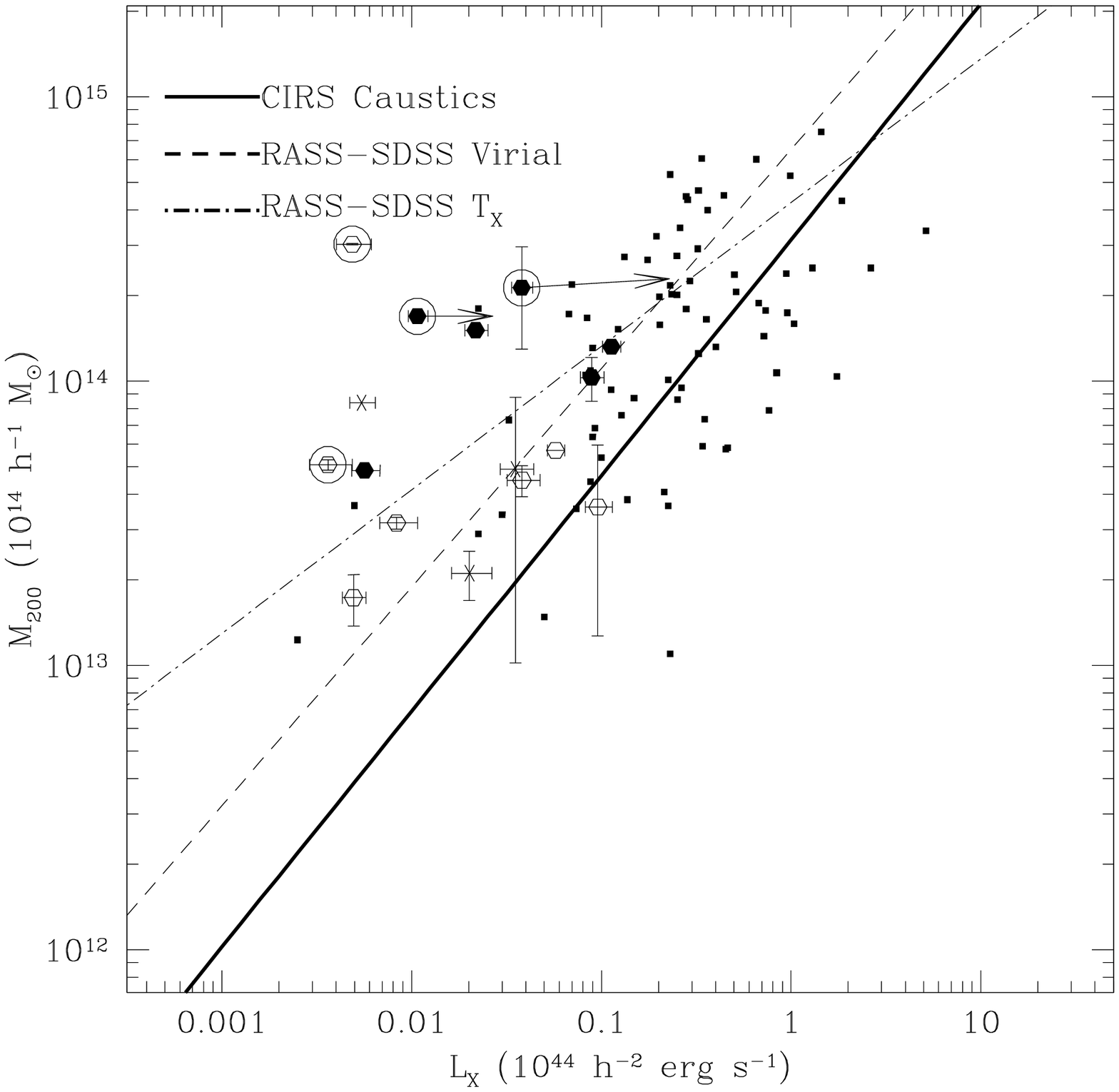}
\caption{\label{clx} Left: Caustic masses at $r_{500}$ compared to X-ray 
luminosities. The solid line is the bisector of the ordinary least
squares fits to the CIRS data.  The red and blue lines show the
$M_{500}-L_X$ relations for RASS-SDSS \citep{popesso05} for optical
and X-ray masses respectively.  Symbol types are explained in Figure
1; large open circles highlight groups strongly affected by nearby
structures.  The two arrows indicate the locations 
of these data points for the larger clusters located close to two of 
these groups (A1541 near cl1227+0858 and A1783 near cl1343+5546).
Right: Same for $M_{200}-L_X$. }
\end{figure*}

Figure \ref{clx} shows $L_X$ versus $M_{500}$ as estimated from the
caustics for the 400d-SDSS groups.  Figure \ref{clx} also shows the
data from CIRS and the best-fit $M_{500}-L_X$ relations for CIRS
(caustic masses) and RASS-SDSS \citep[both virial masses and
$T_X$-based masses][]{popesso05}.

\begin{figure}
\figurenum{11}
\plotone{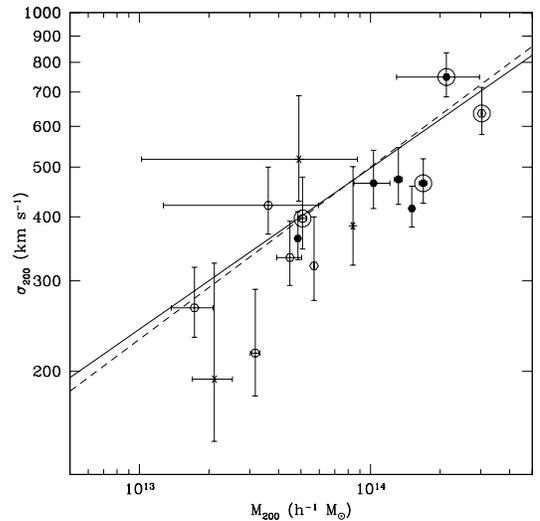} 
\caption{\label{msigma} Caustic masses at $r_{200}$ compared to velocity 
dispersions within $r_{200}$.  The solid line is the bisector of the
ordinary least squares fits for the CIRS clusters.  Symbol types are
explained in Figure 1; large open circles highlight groups strongly
affected by nearby structures.  The dashed line shows the virial mass-$\sigma_p$ relation for dark matter halos in simulations \citep{evrard07}.}
\end{figure}

Figure \ref{msigma} shows the $M_{200} - \sigma _p$ relation.  The
tight relation indicates that the caustic masses are well correlated
with velocity dispersion estimates.  The good correlation is perhaps
not surprising because both parameters depend on the galaxy velocity
distribution.  The best-fit slope for the CIRS clusters is
$M_{200}\propto\sigma_p^{3.18\pm0.19}$ with the uncertainty estimated
from jackknife resampling.  The dashed line in Figure \ref{msigma}
shows the virial mass-$\sigma_p$ relation for dark matter halos in
simulations \citep[][slope 3.035$\pm$0.023]{evrard07}.  The agreement
between these relations from CIRS and simulations further supports the use of
caustics as a mass estimator  (we compare the caustic masses to virial
mass estimates in $\S$\ref{virialsec}).  The four merging groups
in 400d-SDSS represent the most extreme outliers from the CIRS
relation.  The remaining 400d-SDSS groups approximately follow the
CIRS relations, but we do not attempt to fit a relation
for the 400d-SDSS groups.

In the previous section we found that many of the 400d groups are
either group-group mergers (cl0820+5645) or group-cluster mergers
(cl1033+5703, cl1227+0858, cl1343+5546).  These groups represent the
most extreme outliers in the $L_X-\sigma_p$ relation (Figure
\ref{lxsigma}) and the $L_X-M$ relations (Figure \ref{clx}).  This
result reinforces the idea that a detailed understanding of the
large-scale structure surrounding low-mass systems is critical to
obtaining accurate scaling relations.  Previous studies,
\citep[e.g.,][]{helsdon00b,rasscals} usually exclude such systems for
similar reasons.

\section{Discussion}

We use the Fifth Data Release of the Sloan Digital Sky Survey to study
galaxy groups and compare their properties to more massive clusters.
In particular,we study the presence of infall patterns around galaxy
groups, use these patterns to measure mass profiles, and use these
masses to compare the scaling relations of groups and clusters.  This
study extends our previous work on infall patterns around X-ray
clusters in SDSS \citep{cirsi} to smaller X-ray fluxes more typical of
groups.  These systems typically have smaller X-ray luminosities and
masses than the systems studied in previous investigations.  Despite
their smaller masses (compared to CAIRNS and CIRS clusters),
well-defined infall patterns are present in most of these groups.
Four of the 16 groups are contaminated by large-scale structure,
preventing an accurate determination of their infall patterns.

We use the infall patterns to compute mass profiles for the groups and
compare them to model profiles.  Group infall regions are well fit by
NFW and Hernquist profiles and poorly fit by singular isothermal
spheres, similar to the results from CAIRNS and CIRS for clusters.
Observed clusters and groups resemble those in simulations, and their
mass profiles are well described by extrapolations of NFW or Hernquist
models out to the turnaround radius.  The shapes of the best-fit NFW
group mass profiles agree reasonably well with the predictions of
simulations; the average concentration is $c_{200}$$\approx$6.5,
slightly smaller than the value for group size halos extrapolated from
\citet{bullock01} and with similar scatter.  These mass profiles test
the shapes of dark matter haloes on a scale difficult to probe with
weak lensing or any other mass estimator.

There has been much discussion of whether galaxy groups follow
different scaling relations than clusters, particularly with respect
to the $L_X-\sigma_p$ relation 
\citep[e.g.,][]{rasscals,helsdon00b,osmond04}.  Using the caustics to
determine group membership, velocity dispersions, and mass profiles,
the 400d-SDSS groups generally follow the same scaling relations as
clusters, but the small number of 400d-SDSS groups does not allow us
to exclude results suggesting that groups have a shallower
$L_X-\sigma_p$ relation.

Understanding scaling relations of clusters and groups is critical for
future attempts to constrain dark energy with cluster abundance
measurements.  Detailed mass estimates are difficult to obtain for
groups and low-mass clusters because their X-ray emission is typically
faint and because they contain fewer galaxies than massive clusters.
This work represents part of an effort to increase the sample of
groups and low-mass clusters with detailed mass estimates.  Future
work is needed to measure cluster scaling relations robustly across a
wide mass range.  In particular, deep high-resolution X-ray
observations of groups like those studied here could enable
measurements of group masses and X-ray temperatures, while additional
optical spectroscopy could improve measurements of velocity
dispersions and dynamical masses.

\acknowledgements

We thank the referee for a careful reading of the manuscript and for
several helpful suggestions that significantly improved the
presentation of the results.  We thank Margaret Geller for useful
discussions during early stages of this work and for insightful
comments on early drafts.  Support from the PRIN2006 grant
``Costituenti fondamentali dell'Universo'' of the Italian Ministry of
University and Scientific Research and from the INFN grant PD51 is
gratefully acknowledged.  Funding for the Sloan Digital Sky Survey
(SDSS) has been provided by the Alfred P. Sloan Foundation, the
Participating Institutions, the National Aeronautics and Space
Administration, the National Science Foundation, the U.S. Department
of Energy, the Japanese Monbukagakusho, and the Max Planck
Society. The SDSS Web site is http://www.sdss.org/.  The SDSS is
managed by the Astrophysical Research Consortium (ARC) for the
Participating Institutions. The Participating Institutions are The
University of Chicago, Fermilab, the Institute for Advanced Study, the
Japan Participation Group, The Johns Hopkins University, the Korean
Scientist Group, Los Alamos National Laboratory, the
Max-Planck-Institute for Astronomy (MPIA), the Max-Planck-Institute
for Astrophysics (MPA), New Mexico State University, University of
Pittsburgh, University of Portsmouth, Princeton University, the United
States Naval Observatory, and the University of Washington.

\bibliographystyle{apj}
\bibliography{rines}

\begin{thebibliography}{88}
\expandafter\ifx\csname natexlab\endcsname\relax\def\natexlab#1{#1}\fi

\bibitem[{{Abell} {et~al.}(1989){Abell}, {Corwin}, \& {Olowin}}]{aco1989}
{Abell}, G.~O., {Corwin}, H.~G., \& {Olowin}, R.~P. 1989, \apjs, 70, 1

\bibitem[{{Adelman-McCarthy} {et~al.}(2007)}]{dr5}
{Adelman-McCarthy}, J. {et~al.} 2007, \apjs, 172, 634

\bibitem[{{B{\" o}hringer} {et~al.}(2000{\natexlab{a}})}]{2000ApJS..129..435B}
{B{\" o}hringer}, H. {et~al.} 2000{\natexlab{a}}, \apjs, 129, 435

\bibitem[{{B{\" o}hringer} {et~al.}(2000{\natexlab{b}})}]{noras}
---. 2000{\natexlab{b}}, \apjs, 129, 435

\bibitem[{{Biviano} \& {Girardi}(2003)}]{bg03}
{Biviano}, A. \& {Girardi}, M. 2003, \apj, 585, 205

\bibitem[{{Brown} {et~al.}(2009){Brown}, {Geller}, {Kenyon}, \&
  {Diaferio}}]{brown09}
{Brown}, W.~R., {Geller}, M.~J., {Kenyon}, S.~J., \& {Diaferio}, A. 2009,
  submitted to \aj

\bibitem[{{Bullock} {et~al.}(2001)}]{bullock01}
{Bullock}, J.~S. {et~al.} 2001, \mnras, 321, 559

\bibitem[{{Buote} {et~al.}(2007){Buote}, {Gastaldello}, {Humphrey},
  {Zappacosta}, {Bullock}, {Brighenti}, \& {Mathews}}]{buote07}
{Buote}, D.~A., {Gastaldello}, F., {Humphrey}, P.~J., {Zappacosta}, L.,
  {Bullock}, J.~S., {Brighenti}, F., \& {Mathews}, W.~G. 2007, \apj, 664, 123

\bibitem[{{Burenin} {et~al.}(2007){Burenin}, {Vikhlinin}, {Hornstrup},
  {Ebeling}, {Quintana}, \& {Mescheryakov}}]{burenin07}
{Burenin}, R.~A., {Vikhlinin}, A., {Hornstrup}, A., {Ebeling}, H., {Quintana},
  H., \& {Mescheryakov}, A. 2007, \apjs, 172, 561

\bibitem[{{Burke} {et~al.}(2003){Burke}, {Collins}, {Sharples}, {Romer}, \&
  {Nichol}}]{burke03}
{Burke}, D.~J., {Collins}, C.~A., {Sharples}, R.~M., {Romer}, A.~K., \&
  {Nichol}, R.~C. 2003, \mnras, 341, 1093

\bibitem[{{Busha} {et~al.}(2003){Busha}, {Adams}, {Wechsler}, \&
  {Evrard}}]{busha03}
{Busha}, M.~T., {Adams}, F.~C., {Wechsler}, R.~H., \& {Evrard}, A.~E. 2003,
  \apj, 596, 713

\bibitem[{{Busha} {et~al.}(2005){Busha}, {Evrard}, {Adams}, \&
  {Wechsler}}]{busha05}
{Busha}, M.~T., {Evrard}, A.~E., {Adams}, F.~C., \& {Wechsler}, R.~H. 2005,
  \mnras, 363, L11

\bibitem[{{Carballo} {et~al.}(1995){Carballo}, {Warwick}, {Barcons},
  {Gonzalez-Serrano}, {Barber}, {Martinez-Gonzalez}, {Perez-Fournon}, \&
  {Burgos}}]{1995MNRAS.277.1312C}
{Carballo}, R., {Warwick}, R.~S., {Barcons}, X., {Gonzalez-Serrano}, J.~I.,
  {Barber}, C.~R., {Martinez-Gonzalez}, E., {Perez-Fournon}, I., \& {Burgos},
  J. 1995, \mnras, 277, 1312

\bibitem[{{Carlberg} {et~al.}(1997){Carlberg}, {Yee}, \& {Ellingson}}]{cye97}
{Carlberg}, R.~G., {Yee}, H.~K.~C., \& {Ellingson}, E. 1997, \apj, 478, 462

\bibitem[{{Clowe} \& {Schneider}(2001)}]{clowe01}
{Clowe}, D. \& {Schneider}, P. 2001, \aap, 379, 384

\bibitem[{{Comerford} \& {Natarajan}(2007)}]{comerford07}
{Comerford}, J.~M. \& {Natarajan}, P. 2007, \mnras, 379, 190

\bibitem[{{Cross} {et~al.}(2004){Cross}, {Driver}, {Liske}, {Lemon}, {Peacock},
  {Cole}, {Norberg}, \& {Sutherland}}]{2004MNRAS.349..576C}
{Cross}, N.~J.~G., {Driver}, S.~P., {Liske}, J., {Lemon}, D.~J., {Peacock},
  J.~A., {Cole}, S., {Norberg}, P., \& {Sutherland}, W.~J. 2004, \mnras, 349,
  576

\bibitem[{{Dai} {et~al.}(2007){Dai}, {Kochanek}, \& {Morgan}}]{dai07}
{Dai}, X., {Kochanek}, C.~S., \& {Morgan}, N.~D. 2007, \apj, 658, 917

\bibitem[{{Danese} {et~al.}(1980){Danese}, {de Zotti}, \& {di Tullio}}]{danese}
{Danese}, L., {de Zotti}, G., \& {di Tullio}, G. 1980, \aap, 82, 322

\bibitem[{{Diaferio}(1999)}]{diaferio1999}
{Diaferio}, A. 1999, \mnras, 309, 610

\bibitem[{{Diaferio}(2009)}]{diaferio09}
---. 2009, ArXiv e-prints

\bibitem[{{Diaferio} \& {Geller}(1997)}]{dg97}
{Diaferio}, A. \& {Geller}, M.~J. 1997, \apj, 481, 633

\bibitem[{{Diaferio} {et~al.}(2005){Diaferio}, {Geller}, \&
  {Rines}}]{diaferio05}
{Diaferio}, A., {Geller}, M.~J., \& {Rines}, K.~J. 2005, \apjl, 628, L97

\bibitem[{{Drinkwater} {et~al.}(2001){Drinkwater}, {Gregg}, \&
  {Colless}}]{drink}
{Drinkwater}, M.~J., {Gregg}, M.~D., \& {Colless}, M. 2001, \apj, 548, L139

\bibitem[{{Ebeling} {et~al.}(1998){Ebeling}, {Edge}, {Bohringer}, {Allen},
  {Crawford}, {Fabian}, {Voges}, \& {Huchra}}]{bcs}
{Ebeling}, H., {Edge}, A.~C., {Bohringer}, H., {Allen}, S.~W., {Crawford},
  C.~S., {Fabian}, A.~C., {Voges}, W., \& {Huchra}, J.~P. 1998, \mnras, 301,
  881

\bibitem[{{Eke} {et~al.}(1996){Eke}, {Cole}, \& {Frenk}}]{ecf96}
{Eke}, V.~R., {Cole}, S., \& {Frenk}, C.~S. 1996, \mnras, 282, 263

\bibitem[{{Evrard} {et~al.}(2008)}]{evrard07}
{Evrard}, A.~E. {et~al.} 2008, \apj, 672, 122

\bibitem[{{Falco} {et~al.}(1999)}]{falco99}
{Falco}, E.~E. {et~al.} 1999, \pasp, 111, 438

\bibitem[{{Geller} {et~al.}(1999){Geller}, {Diaferio}, \& {Kurtz}}]{gdk99}
{Geller}, M.~J., {Diaferio}, A., \& {Kurtz}, M.~J. 1999, \apjl, 517, L23

\bibitem[{{Girardi} {et~al.}(1998){Girardi}, {Giuricin}, {Mardirossian},
  {Mezzetti}, \& {Boschin}}]{girardi98}
{Girardi}, M., {Giuricin}, G., {Mardirossian}, F., {Mezzetti}, M., \&
  {Boschin}, W. 1998, \apj, 505, 74

\bibitem[{{Gramann} \& {Suhhonenko}(2002)}]{2002MNRAS.337.1417G}
{Gramann}, M. \& {Suhhonenko}, I. 2002, \mnras, 337, 1417

\bibitem[{{Heisler} {et~al.}(1985){Heisler}, {Tremaine}, \& {Bahcall}}]{htb}
{Heisler}, J., {Tremaine}, S., \& {Bahcall}, J.~N. 1985, \apj, 298, 8

\bibitem[{{Helsdon} \& {Ponman}(2000)}]{helsdon00b}
{Helsdon}, S.~F. \& {Ponman}, T.~J. 2000, \mnras, 319, 933

\bibitem[{{Helsdon} {et~al.}(2005){Helsdon}, {Ponman}, \&
  {Mulchaey}}]{helsdon05}
{Helsdon}, S.~F., {Ponman}, T.~J., \& {Mulchaey}, J.~S. 2005, \apj, 618, 679

\bibitem[{{Hernquist}(1990)}]{hernquist1990}
{Hernquist}, L. 1990, \apj, 356, 359

\bibitem[{{Kaiser}(1987)}]{kais87}
{Kaiser}, N. 1987, \mnras, 227, 1

\bibitem[{{Khosroshahi} {et~al.}(2007){Khosroshahi}, {Ponman}, \&
  {Jones}}]{khosroshahi07}
{Khosroshahi}, H.~G., {Ponman}, T.~J., \& {Jones}, L.~R. 2007, \mnras, 377, 595

\bibitem[{{Kneib} {et~al.}(2003)}]{kneib03}
{Kneib}, J.-P. {et~al.} 2003, \apj, 598, 804

\bibitem[{{Koranyi} \& {Geller}(2000)}]{kg2000}
{Koranyi}, D.~M. \& {Geller}, M.~J. 2000, \aj, 119, 44

\bibitem[{{Ledlow} {et~al.}(2003){Ledlow}, {Voges}, {Owen}, \&
  {Burns}}]{2003AJ....126.2740L}
{Ledlow}, M.~J., {Voges}, W., {Owen}, F.~N., \& {Burns}, J.~O. 2003, \aj, 126,
  2740

\bibitem[{{Lemze} {et~al.}(2009){Lemze}, {Broadhurst}, {Rephaeli}, {Barkana},
  \& {Umetsu}}]{lemze09}
{Lemze}, D., {Broadhurst}, T., {Rephaeli}, Y., {Barkana}, R., \& {Umetsu}, K.
  2009, \apj, 701, 1336

\bibitem[{{Lin} {et~al.}(2004){Lin}, {Mohr}, \& {Stanford}}]{lin04}
{Lin}, Y., {Mohr}, J.~J., \& {Stanford}, S.~A. 2004, \apj, 610, 745

\bibitem[{{{\L}okas} \& {Mamon}(2003)}]{lokas03}
{{\L}okas}, E.~L. \& {Mamon}, G.~A. 2003, \mnras, 343, 401

\bibitem[{{Lopes} {et~al.}(2009){Lopes}, {de Carvalho}, {Kohl-Moreira}, \&
  {Jones}}]{lopes09}
{Lopes}, P.~A.~A., {de Carvalho}, R.~R., {Kohl-Moreira}, J.~L., \& {Jones}, C.
  2009, \mnras, 392, 135

\bibitem[{{Mahdavi} {et~al.}(2000){Mahdavi}, {B{\" o}hringer}, {Geller}, \&
  {Ramella}}]{rasscals}
{Mahdavi}, A., {B{\" o}hringer}, H., {Geller}, M.~J., \& {Ramella}, M. 2000,
  \apj, 534, 114

\bibitem[{{Mahdavi} \& {Geller}(2001)}]{andilxsig}
{Mahdavi}, A. \& {Geller}, M.~J. 2001, \apjl, 554, L129

\bibitem[{{Mahdavi} {et~al.}(1999){Mahdavi}, {Geller}, {B{\" o}hringer},
  {Kurtz}, \& {Ramella}}]{mahdavi99}
{Mahdavi}, A., {Geller}, M.~J., {B{\" o}hringer}, H., {Kurtz}, M.~J., \&
  {Ramella}, M. 1999, \apj, 518, 69

\bibitem[{{Mahdavi} {et~al.}(2005){Mahdavi}, {Trentham}, \&
  {Tully}}]{mahdavi05}
{Mahdavi}, A., {Trentham}, N., \& {Tully}, R.~B. 2005, \aj, 130, 1502

\bibitem[{{Majumdar} \& {Mohr}(2004)}]{majumdar04}
{Majumdar}, S. \& {Mohr}, J.~J. 2004, \apj, 613, 41

\bibitem[{{Mandelbaum} {et~al.}(2006){Mandelbaum}, {Seljak}, {Cool}, {Blanton},
  {Hirata}, \& {Brinkmann}}]{mandelbaum06}
{Mandelbaum}, R., {Seljak}, U., {Cool}, R.~J., {Blanton}, M., {Hirata}, C.~M.,
  \& {Brinkmann}, J. 2006, \mnras, 372, 758

\bibitem[{{Mantz} {et~al.}(2008){Mantz}, {Allen}, {Ebeling}, \&
  {Rapetti}}]{mantz08}
{Mantz}, A., {Allen}, S.~W., {Ebeling}, H., \& {Rapetti}, D. 2008, \mnras, 387,
  1179

\bibitem[{{Masson}(1979)}]{1979MNRAS.187..253M}
{Masson}, C.~R. 1979, \mnras, 187, 253

\bibitem[{{Merch{\'a}n} \& {Zandivarez}(2005)}]{merchan05}
{Merch{\'a}n}, M.~E. \& {Zandivarez}, A. 2005, \apj, 630, 759

\bibitem[{{Miller} {et~al.}(2005)}]{miller05}
{Miller}, C.~J. {et~al.} 2005, \aj, 130, 968

\bibitem[{{Moore} {et~al.}(1999){Moore}, {Quinn}, {Governato}, {Stadel}, \&
  {Lake}}]{moore99}
{Moore}, B., {Quinn}, T., {Governato}, F., {Stadel}, J., \& {Lake}, G. 1999,
  \mnras, 310, 1147

\bibitem[{{Mulchaey} \& {Zabludoff}(1998)}]{mulchaey98}
{Mulchaey}, J.~S. \& {Zabludoff}, A.~I. 1998, \apj, 496, 73

\bibitem[{{Nagamine} \& {Loeb}(2003)}]{nl02}
{Nagamine}, K. \& {Loeb}, A. 2003, New Astronomy, 8, 439

\bibitem[{{Navarro} {et~al.}(1997){Navarro}, {Frenk}, \& {White}}]{nfw97}
{Navarro}, J.~F., {Frenk}, C.~S., \& {White}, S. D.~M. 1997, \apj, 490, 493

\bibitem[{{Osmond} \& {Ponman}(2004)}]{osmond04}
{Osmond}, J.~P.~F. \& {Ponman}, T.~J. 2004, \mnras, 350, 1511

\bibitem[{{Osmond} {et~al.}(2004){Osmond}, {Ponman}, \&
  {Finoguenov}}]{osmond04b}
{Osmond}, J.~P.~F., {Ponman}, T.~J., \& {Finoguenov}, A. 2004, \mnras, 355, 11

\bibitem[{{Ponman} {et~al.}(1996){Ponman}, {Bourner}, {Ebeling}, \&
  {B{\"o}hringer}}]{ponman96}
{Ponman}, T.~J., {Bourner}, P.~D.~J., {Ebeling}, H., \& {B{\"o}hringer}, H.
  1996, \mnras, 283, 690

\bibitem[{{Popesso} {et~al.}(2007){Popesso}, {Biviano}, {B{\"o}hringer}, \&
  {Romaniello}}]{popesso06}
{Popesso}, P., {Biviano}, A., {B{\"o}hringer}, H., \& {Romaniello}, M. 2007,
  \aap, 461, 397

\bibitem[{{Popesso} {et~al.}(2005){Popesso}, {Biviano}, {B{\"o}hringer},
  {Romaniello}, \& {Voges}}]{popesso05}
{Popesso}, P., {Biviano}, A., {B{\"o}hringer}, H., {Romaniello}, M., \&
  {Voges}, W. 2005, \aap, 433, 431

\bibitem[{{Popesso} {et~al.}(2004){Popesso}, {B{\"o}hringer}, {Brinkmann},
  {Voges}, \& {York}}]{popesso04}
{Popesso}, P., {B{\"o}hringer}, H., {Brinkmann}, J., {Voges}, W., \& {York},
  D.~G. 2004, \aap, 423, 449

\bibitem[{{Quintana} \& {Melnick}(1982)}]{quintana82}
{Quintana}, H. \& {Melnick}, J. 1982, \aj, 87, 972

\bibitem[{{Rasmussen} {et~al.}(2006){Rasmussen}, {Ponman}, {Mulchaey}, {Miles},
  \& {Raychaudhury}}]{rasmussen06}
{Rasmussen}, J., {Ponman}, T.~J., {Mulchaey}, J.~S., {Miles}, T.~A., \&
  {Raychaudhury}, S. 2006, \mnras, 373, 653

\bibitem[{{Reg\"os} \& {Geller}(1989)}]{rg89}
{Reg\"os}, E. \& {Geller}, M.~J. 1989, \aj, 98, 755

\bibitem[{{Reisenegger} {et~al.}(2000){Reisenegger}, {Quintana}, {Carrasco}, \&
  {Maze}}]{rqcm}
{Reisenegger}, A., {Quintana}, H., {Carrasco}, E.~R., \& {Maze}, J. 2000, \aj,
  120, 523

\bibitem[{{Rines} {et~al.}(2002){Rines}, {Geller}, {Diaferio}, {Mahdavi},
  {Mohr}, \& {Wegner}}]{rines02}
{Rines}, K., {Geller}, M.~J., {Diaferio}, A., {Mahdavi}, A., {Mohr}, J.~J., \&
  {Wegner}, G. 2002, \aj, 124, 1266

\bibitem[{{Rines} {et~al.}(2000){Rines}, {Geller}, {Diaferio}, {Mohr}, \&
  {Wegner}}]{rines2000}
{Rines}, K., {Geller}, M.~J., {Diaferio}, A., {Mohr}, J.~J., \& {Wegner}, G.~A.
  2000, \aj, 120, 2338

\bibitem[{{Rines} {et~al.}(2003){Rines}, {Geller}, {Kurtz}, \&
  {Diaferio}}]{cairnsi}
{Rines}, K., {Geller}, M.~J., {Kurtz}, M.~J., \& {Diaferio}, A. 2003, \aj, 126,
  2152

\bibitem[{{Rines} {et~al.}(2001){Rines}, {Mahdavi}, {Geller}, {Diaferio},
  {Mohr}, \& {Wegner}}]{rines01b}
{Rines}, K., {Mahdavi}, A., {Geller}, M.~J., {Diaferio}, A., {Mohr}, J.~J., \&
  {Wegner}, G. 2001, \apj, 555, 558

\bibitem[{{Rines} \& {Diaferio}(2006)}]{cirsi}
{Rines}, K.~J. \& {Diaferio}, A. 2006, \aj, 132, 1275

\bibitem[{{Rykoff} {et~al.}(2008)}]{rykoff08}
{Rykoff}, E.~S. {et~al.} 2008, \apj, 675, 1106

\bibitem[{{Serra} {et~al.}(2009){Serra}, {Angus}, \& {Diaferio}}]{serra09}
{Serra}, A.~L., {Angus}, G.~W., \& {Diaferio}, A. 2009, ArXiv e-prints

\bibitem[{{Slinglend} {et~al.}(1998){Slinglend}, {Batuski}, {Miller}, {Haase},
  {Michaud}, \& {Hill}}]{1998ApJS..115....1S}
{Slinglend}, K., {Batuski}, D., {Miller}, C., {Haase}, S., {Michaud}, K., \&
  {Hill}, J.~M. 1998, \apjs, 115, 1

\bibitem[{{Small} {et~al.}(1997){Small}, {Sargent}, \& {Hamilton}}]{small97}
{Small}, T.~A., {Sargent}, W. L.~W., \& {Hamilton}, D. 1997, \apjs, 111, 1

\bibitem[{{Smith} {et~al.}(2000){Smith}, {Lucey}, {Hudson}, {Schlegel}, \&
  {Davies}}]{rsmith00}
{Smith}, R.~J., {Lucey}, J.~R., {Hudson}, M.~J., {Schlegel}, D.~J., \&
  {Davies}, R.~L. 2000, \mnras, 313, 469

\bibitem[{{Stanek} {et~al.}(2006){Stanek}, {Evrard}, {B{\"o}hringer},
  {Schuecker}, \& {Nord}}]{stanek06}
{Stanek}, R., {Evrard}, A.~E., {B{\"o}hringer}, H., {Schuecker}, P., \& {Nord},
  B. 2006, \apj, 648, 956

\bibitem[{{Stoughton} {et~al.}(2002)}]{sdss}
{Stoughton}, C. {et~al.} 2002, \aj, 123, 485

\bibitem[{{Strauss} {et~al.}(2002)}]{strauss02}
{Strauss}, M.~A. {et~al.} 2002, \aj, 124, 1810

\bibitem[{{Vedel} \& {Hartwick}(1998)}]{vh98}
{Vedel}, H. \& {Hartwick}, F.~D.~A. 1998, \apj, 501, 509

\bibitem[{{Vikhlinin} {et~al.}(1998){Vikhlinin}, {McNamara}, {Forman}, {Jones},
  {Quintana}, \& {Hornstrup}}]{vikhlinin98}
{Vikhlinin}, A., {McNamara}, B.~R., {Forman}, W., {Jones}, C., {Quintana}, H.,
  \& {Hornstrup}, A. 1998, \apj, 502, 558

\bibitem[{{Vikhlinin} {et~al.}(1999){Vikhlinin}, {McNamara}, {Hornstrup},
  {Quintana}, {Forman}, {Jones}, \& {Way}}]{1999ApJ...520L...1V}
{Vikhlinin}, A., {McNamara}, B.~R., {Hornstrup}, A., {Quintana}, H., {Forman},
  W., {Jones}, C., \& {Way}, M. 1999, \apjl, 520, L1

\bibitem[{{Voevodkin} {et~al.}(2009){Voevodkin}, {Borozdin}, {Heitmann},
  {Habib}, {Vikhlinin}, {Mescheryakov}, \& {Hornstrup}}]{voevodkin09}
{Voevodkin}, A., {Borozdin}, K., {Heitmann}, K., {Habib}, S., {Vikhlinin}, A.,
  {Mescheryakov}, A., \& {Hornstrup}, A. 2009, ArXiv e-prints

\bibitem[{{Voges} {et~al.}(1999)}]{rass}
{Voges}, W. {et~al.} 1999, \aap, 349, 389

\bibitem[{{Wu} {et~al.}(1999){Wu}, {Xue}, \& {Fang}}]{wu99}
{Wu}, X.-P., {Xue}, Y.-J., \& {Fang}, L.-Z. 1999, \apj, 524, 22

\bibitem[{{Xue} \& {Wu}(2000)}]{xue00}
{Xue}, Y.-J. \& {Wu}, X.-P. 2000, \apj, 538, 65

\end{thebibliography}

\clearpage

\end{document}